\documentclass[%
 reprint,
superscriptaddress,
groupedaddress,
 amsmath,amssymb,
 aps,
]{revtex4-2}

\usepackage{booktabs}
\usepackage{amsmath}

\usepackage{graphicx}   
\usepackage[colorlinks=true, allcolors=blue]{hyperref}
\usepackage{dcolumn}    
\usepackage{bm}         
\usepackage{rotating}
\usepackage[table,xcdraw]{xcolor} 

\usepackage{caption}
\usepackage{subcaption}
\usepackage[export]{adjustbox}

\begin{document}

\preprint{APS/123-QED}

\title{Average power-density spectrum \\of short and long Fermi-GBM gamma-ray bursts}

\author{Else Magnus} 
\email{else.magnus@vub.be} 
\affiliation{Department of Physics and Astronomy, Vrije Universiteit Brussel, Elsene, Belgium}
\author{Jannes Loonen} 
\affiliation{Department of Physics and Astronomy, Vrije Universiteit Brussel, Elsene, Belgium}
\author{Rose S. Stanley} 
\affiliation{Department of Physics and Astronomy, Vrije Universiteit Brussel, Elsene, Belgium}
\author{Paul Coppin}
\affiliation{Department of Nuclear and Particle Physics, University of Geneva, CH-1211 Geneva, Switzerland}
\author{Krijn D. de Vries}
\email{krijn.de.vries@vub.be}
\affiliation{Department of Physics and Astronomy, Vrije Universiteit Brussel, Elsene, Belgium}
\author{Nick van Eijndhoven}
\email{nick.van.eijndhoven@vub.be}
\affiliation{Department of Physics and Astronomy, Vrije Universiteit Brussel, Elsene, Belgium}

\date{October 17, 2025}

\begin{abstract}

    Gamma-ray bursts (GRBs) are the most powerful electromagnetic outbursts in the Universe and emit a vast amount of their energy in the form of gamma rays. Their duration is extremely short on cosmic timescales, but they show a wealth of time variability in their light curves. Properties of this variability may carry information about the processes the gamma rays emerge from, which are still poorly understood. 
    This research investigates the redshift-corrected gamma-ray light curves of 159 long GRBs, observed with the Gamma-Ray Burst Monitor on the \textit{Fermi} Gamma-Ray Space Telescope between 2008 and 2023. We calculate the average power-density spectrum of different groups of GRBs that are distinguished based on fluence, peak rate, duration, redshift, and the different GRB phases. 
    Almost all redshift-corrected spectra reveal a power-law behavior with high-frequency power-law indices distributed around $\sim -1.9 \pm 0.2$. The precursor phase and redshift-corrected short bursts exhibit a shallower power law with index $\sim -1.30 \pm 0.04$, potentially due to the limited statistics that these samples represent. Only in some cases is the high-frequency index still consistent with the $-5/3$ (Kolmogorov) slope, found by earlier studies and linked to the appearance of fully developed turbulence. 

\end{abstract}

\maketitle


\section{\label{sec:introduction} INTRODUCTION}

    Gamma-ray bursts (GRBs) are very short, second-to-minute duration, bright flashes of gamma rays with a typical luminosity of $\sim 10^{51}$-$10^{53}$ erg $\mathrm{s^{-1}}$ \cite{Zhang_2018}. The main burst with emission in the hard x-ray or soft $\gamma$-ray band is called the prompt emission and is followed by a broadband afterglow emission that can last for hours to days \cite{Kimura_2022}. Their prompt-duration histogram is bimodal, suggesting a natural classification into two subclasses: short GRBs (SGRBs) with $T_{90} < 2$ s and long GRBs (LGRBs) with $T_{90} > 2$ s \cite{Kouveliotou_1993}. The $T_{90}$ is the time interval during which the observed burst fluence ranges from 5\% to 95\% of the total observed fluence. Typically, long GRBs have a mean $T_{90} \sim 30$ s, and short GRBs have a mean $T_{90} \sim 1$ s \cite{von_Kienlin_2020}. Rarely, in ($3$-$20$)\% of all GRBs, the prompt emission is preceded by precursor emission, a shorter and weaker gamma-ray pulse \cite{Li_2022, Coppin_2020}. The natural selection of GRBs into long and short GRBs is believed to originate from different progenitors, strengthened by multiwavelength and multimessenger observations. Short GRBs are typically connected to compact binary systems \cite{Paczynski_1991, Eichler_1989, Narayan_1991, Abbott_2017}, while long GRBs are observed in combination with supernova features \cite{Woosly_1993, Paczynski_1998, MacFadyen_1999}. However, recent discoveries of peculiar GRBs that do not align with this classification, challenge this hypothesis \cite{Zhu_2024, Zhang_2025}.
    
    The gamma-ray light curves of the various GRBs are seen to be largely different. Some are smooth (i.e. contain well-defined peaks), while the majority of the GRBs contain complex multiepisode patterns \cite{Fishman_1995, Lin_2013}. While these structures contain a wealth of information about the dissipation and radiation mechanisms of the gamma rays, a complete understanding of the GRB temporal variability is still lacking \cite{Camisasca_2023}. The diversity of light curves implies that complicated mechanisms lie at the origin of the GRB radiation \cite{Tarnopolski_2021}. Understanding the temporal variability can help to gain insight in the underlying mechanisms of the GRB \cite{Guidorzi_2012, Dichiara_2013a, MacLachlan_2013, Guidorzi_2016, Dichiara_2016, Kumar_2015, Zhou_2024, Guidorzi_2024}. Indeed, several questions remain unanswered. What are the physical dissipation mechanisms that produce the prompt emission and early afterglow emission: internal or external shocks or something different? Are they dominantly produced by either synchrotron or inverse-Compton processes? At which radius from the central engine are the gamma rays produced? Are the ejected jets primarily baryon dominated or magnetic dominated \cite{Kumar_2015, Zhang_2018, Piran_2004}? 
    
    Temporal characteristics are often studied through the power-density spectrum (PDS), being the square of the Fourier-transform amplitude. Extensive research has been performed during the past 25 years to resolve the dominant timescales in GRB gamma-ray light curves. Belobodorov, Stern, and Svensson applied a Fourier analysis to 214 bright and long GRBs observed by \textit{CGRO}-BATSE (Burst and Transient Source Experiment \cite{Paciesas_1999}) \cite{Beloborodov_1998}. Under the assumption that the same underlying processes occur in all GRBs, they justified the averaging of the individual power-density spectra to construct the average PDS of a group of GRBs. This reduces the random fluctuations due to the diversity of the spectra. They reported a power-law index of the average PDS, consistent with $-5/3$ within uncertainties, which they attributed to the appearance of fully developed turbulence (so-called Kolmogorov turbulence \cite{Kolmogorov_1991, Pope_2000}). Additionally, they noticed a break in the spectrum at a frequency of $f \sim 1$ Hz. These findings were confirmed by their extended follow-up research \cite{Beloborodov_2000} and an investigation of 10 long and bright \textit{INTEGRAL}-SPI GRBs \cite{Vedrenne_2003, Winkler_2003, Ryde_2003}. Subsequently, Lazzati was the first to fit the average PDS with a broken power-law model \cite{Lazzati_2002}. The results suggested a relationship between the break frequency and the degree of variability in the light curves, $V$ \cite{Fenimore_2000}, while the low-frequency power-law index $\beta_{LF}$ and high-frequency power-law index $\beta_{HF}$ remained constant for varying values of $V$ ($\beta_{LF} \sim -2/3$ and $\beta_{HF} \sim -2$). Shen and Song investigated two classes of long bursts in order to distinguish progenitors \cite{Shen_2003}. They found power-law indices of $-1.84$ and $-1.78$ and no significant difference. Borgonovo \textit{et al.}\ divided 22 long \textit{CGRO}-BATSE, \textit{Wind}-Konus \cite{Aptekar_1995}, and \textit{BeppoSAX}-GRBM (Gamma-Ray Burst Monitor \cite{Frontera_1997}) GRBs with known redshift in two classes based on the values of the autocorrelation function (see Ref. \cite{Borgonovo_2007} for a detailed explanation). The resulting average PDS differed for both classes: One was modeled by a single power law with index $-1.97 \pm 0.04$, while the other was better characterized by a combination of a low-frequency stretched exponential and high-frequency power law with index $-1.6 \pm 0.2$, again consistent with the Kolmogorov index.
    
    Guidorzi \textit{et al.}\ investigated the average PDS of 244 long \textit{Swift}-BAT (Burst Alert Telescope \cite{Barthelmy_2005}) GRBs and divided them in different subgroups \cite{Guidorzi_2012}. A broken power law was fitted to the different spectra, and low-frequency indices were found around $\sim - 1.0$, while high-frequency indices varied between $-1.7$ and $-2.0$, without evidence for a break around $f \sim 1$ Hz. In most cases, the high-frequency power-law index was consistent with the Kolmogorov index of $-5/3$. The break frequency $f_b$ ranged between $0.018$ and $0.1$ Hz. This frequency can be related to a characteristic timescale $\tau \sim 1/2\pi f_b = 2$-$4 \ \mathrm{s}$, as follows from the shot-noise model developed by Lazatti \cite{Lazzati_2002}. Additionally, the power-law indices showed a flattening for more energetic bursts and bursts at higher redshifts. Of those 244 GRBs, 97 were used to study the effect of redshift correction. No significant differences in the power-law indices were found between the observer and source frame \cite{Guidorzi_2012}. A follow-up study by Dichiara \textit{et al.}\ on long and bright GRBs, detected by BeppoSAX-GRBM and \textit{Fermi}-GBM (Gamma-ray Burst Monitor \cite{Meegan_2009}), confirmed the energy dependence of the PDS slope \cite{Dichiara_2013a}. More energetic light curves have flatter PDSs. A clear break at $f \sim 1$-$2$ Hz was observed. The low-frequency power-law indices had values around $-1.0$, and the high-frequency power-law indices varied between $-1.5$ and $-2.5$, depending on the sample of GRBs, therefore consistent with $-5/3$. Recent research of Zhou \textit{et al.}\ investigated the average PDS of 64 long GRBs and 11 short GRBs detected by Insight-HXMT-HE (High Energy X-ray telescope \cite{Liu_2020}) \cite{Zhou_2024}. They found a high-frequency power-law index decreasing with energy for both long and short bursts, consistent with the slope-energy relation of Dichiara \textit{et al.}. 

    Finally, the power-law behavior is observed in the individual power-density spectra for short and long GRBs as well. We refer the reader to Refs. \cite{Belli_1992, Beloborodov_1998, Chang_2001, Chang_2002, Ukwatta_2011, Dichiara_2013b, Dichiara_2016, Guidorzi_2016, Zhou_2024}. The power-law slopes of the individual spectra typically range between $\sim -1.0$ and $\sim -4$. Notably, Guidorzi \textit{et al.}\ compare the power-density spectra of GRBs to the PDS properties of other sources that are powered by black-hole accretion, suggesting an analogy between the stochastic accretion processes in GRBs and active galactic nuclei \cite{Guidorzi_2016}. 

    The persistence of the power-law behavior across different energy bands, instrument sensitivities, and other variables strengthens the hypothesis that scale-free processes lie at the origin of the gamma-ray variability. This strongly suggests that each GRB is indeed a stochastic result of the same underlying mechanism. The slope $-5/3$ of the Kolmogorov spectrum resonates in the different searches, although the unique circumstances of the GRB seem to influence this slope as well \cite{Tarnopolski_2021}. In this paper, we investigate the redshift-corrected average PDS of 214 GRBs with known redshift, detected by the \textit{Fermi}-GBM. The structure of the paper is as follows. The selection and analysis of the data are described in Sec. \ref{sec:dataanalysis}. Section \ref{sec:results} reports the results per subgroup, which are subsequently discussed in Sec. \ref{sec:discussion}. Appendixes \ref{app:backgroundcharacterisation} and \ref{app:phaseidentification} contain details about the processing of the GRB light curves.

\section{\label{sec:dataanalysis} DATA ANALYSIS}

\subsection{\label{subsec:dataselection} Data selection}

    In this study, we analyze GRBs observed by the \textit{Fermi} Gamma-Ray Space Telescope using the Gamma-ray Burst Monitor (GBM). The \textit{Fermi}-GBM instrument consists of 12 sodium iodide (NaI) and two bismuth germanate (BGO) detectors. The NaI detectors, covering the 8 keV to 1 MeV energy range, are responsible for detecting and localizing the GRBs by comparing the measured relative count rates across the detectors \cite{Meegan_2009}. The bismuth germanate detectors, which are sensitive to energies from 200 keV to 40 MeV, are not directly used in this study but provide cross-calibration between \textit{Fermi}-GBM and the higher-energy instrument on board, \textit{Fermi}-LAT. 
    
    Our dataset comprises 3709 GRBs detected by the NaI detectors between July 2008 and December 2023. Of these, nine lack the time-tagged event (TTE) data files required for our analysis. We focus on GRBs with known redshift, resulting in a sample of 215 GRBs. One of these objects, GRB 221009553, was reported to have significant TTE data loss due to the high intensity of the burst and the limited bandwidth of the instrument \cite{website:fermi-support}. Given that our study involves the analysis of the time variability of the GRB light curves, we exclude this GRB from the sample. Eight other GRBs experienced short gaps in the TTE data that do not affect the emission zones. They are retained in the analysis. As will be explained in Sec. \ref{subsec:dataprocessing}, ten GRBs will later be removed from the sample due to the absence of identifiable emission phases. The final sample, therefore, consists of 204 GRBs. 
    
\begin{figure*}
        \centering
        \includegraphics[width=0.95\linewidth]{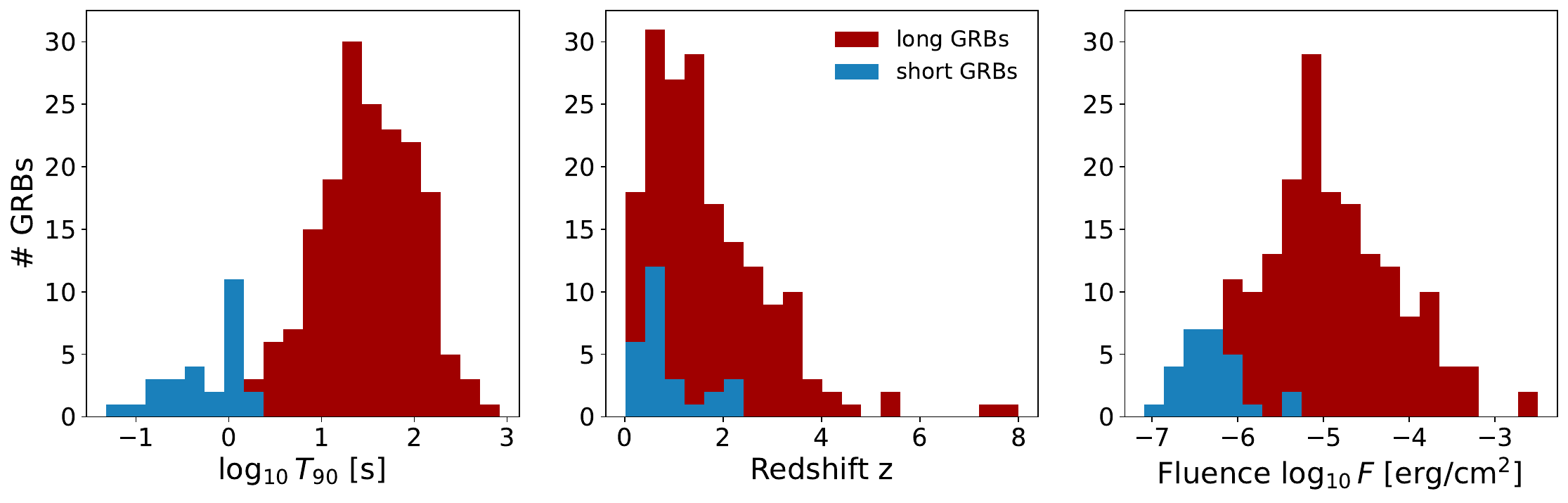}
        \caption{Observed $T_{90}$ distribution (left), redshift distribution (center) and fluence distribution (right) of the 204 selected GRBs, distinguishing between short (blue) and long (red) GRBs.}
        \label{fig:grbdistribution}
    \end{figure*}

    Figure \ref{fig:grbdistribution} displays the observed $T_{90}$ distribution, redshift distribution, and fluence distribution for the selected GRBs. All values are retrieved from the GRBWeb catalog \cite{Coppin_2020}. The $T_{90}$ values range from 0.048 to 828.672 s. The conventional classification of Ref. \citep{Kouveliotou_1993} distinguishes between short and long bursts at $T_{90} = 2$ s. The sample includes 27 SGRBs and 177 LGRBs. The redshift distribution spans from a minimum of $z = 0.0093$ to a maximum of $z = 8.0$. Fluences vary across 6 orders of magnitude, with the lowest fluence recorded at $F = 8.2 \cdot 10^{-8}$ erg/$\mathrm{cm^2}$ and the highest at $F = 3.1 \cdot 10^{-3}$ erg/$\mathrm{cm^2}$.

\subsection{\label{subsec:dataprocessing} Data processing}
    
    The TTE data \cite{fermidata} provided by the NaI detectors stores observed photon counts with a temporal resolution of 2 $\mu$s \cite{Paciesas_2012}. Each detected photon is assigned an energy index corresponding to one of the 128 bins spanning the full energy range of the NaI detectors \cite{Meegan_2009}. Consequently, the assigned energy value does not necessarily represent the photon's true incoming energy. For each GRB, we require that at least two NaI detectors have triggered on the burst, and we select those triggered detectors for analysis. If more than three detectors are triggered, we select the three with the closest alignment to the GRB's direction. The photon counts from the selected detectors are then integrated to construct a combined light curve for each GRB. This light curve will serve as a basis for the subsequent analysis. 

    A redshift correction is applied to the photon arrival times relative to the trigger time and energies using the transformations $t_e = (1+z)^{-1} \cdot t_o$ and $E_e = (1+z) \cdot E_o$. The subscript ``$e$" represents the emitted energy and time in the source frame, and ``$o$" denotes the observer frame. The photon counts are then binned in time using a fixed bin width of $5 \ \mathrm{ms}$, following Ref. \cite{Coppin_2020}. To account for background noise, a background-characterization procedure is performed on each GRB histogram. This procedure relies on the method described in Ref. \cite{Coppin_2020} and is briefly summarized in Appendix \ref{app:backgroundcharacterisation}. It involves identifying and subtracting the background rate from the light curve.  

    \begin{figure}
        \centering
        \includegraphics[width=0.75\linewidth]{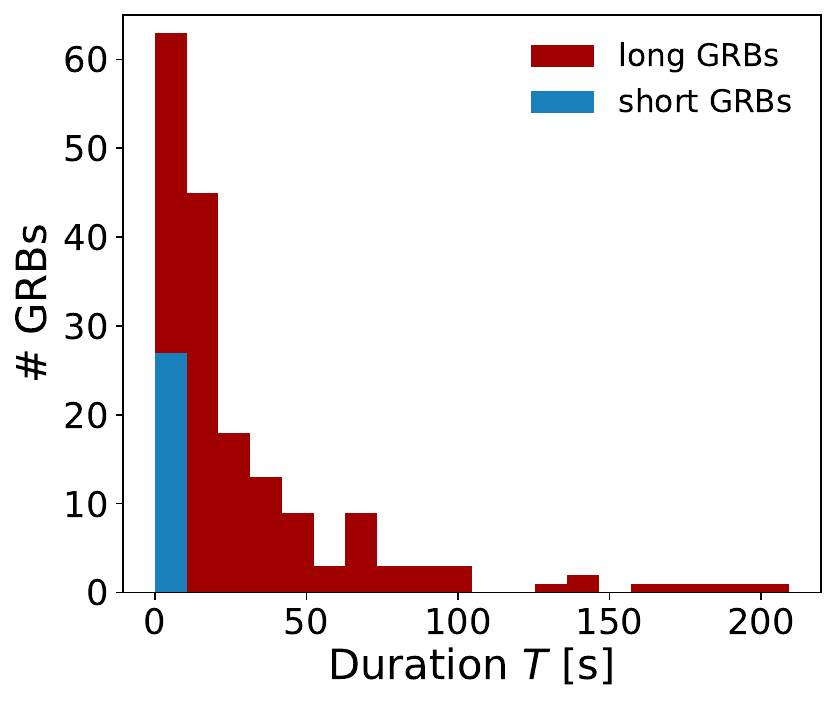}
        \caption{Observed $T$ distribution of the 204 selected GRBs, where $T$ is defined as the duration of the total signal region, identified by the phase identification used in this analysis.}
        \label{fig:grb_distribution_T}
    \end{figure}

    \begin{figure*}
        \centering
        \includegraphics[width=0.95\linewidth]{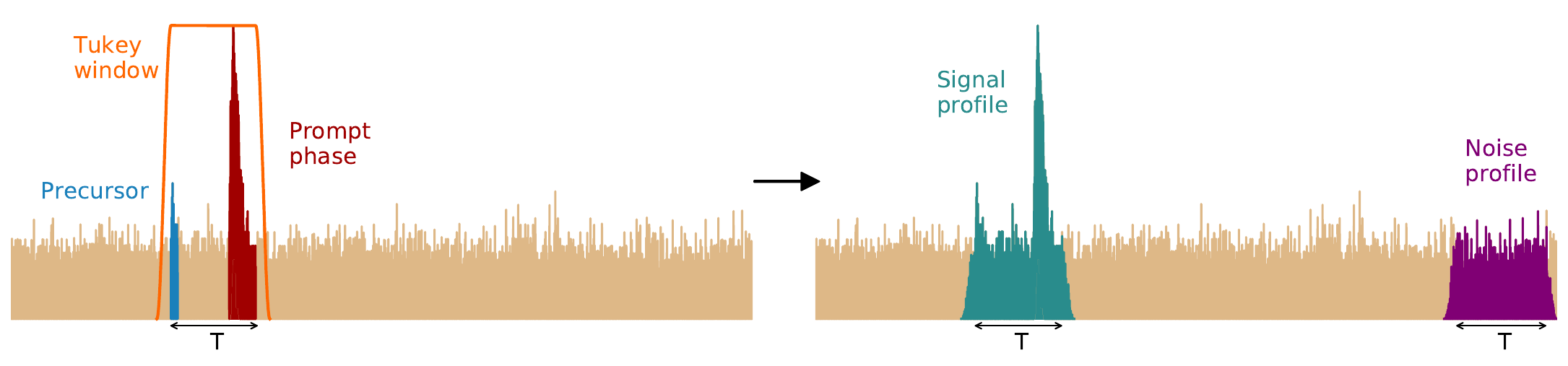}
        \caption{Left: The phase-identification procedure identifies the signal regions in the light curve, in this case a precursor and prompt phase. A Tukey window function with $\alpha = 0.25$ selects only the relevant emission zone in the light curve, based on these signal regions. The duration of the total emission region is called $T$. Right: The PDS analysis is performed on the relevant signal region and, subsequently, on a noise region of the same length.}
        \label{fig:lightcurve_example}
    \end{figure*}
    
    After background subtraction, we identify the signal regions in the light curve, such as the precursor, prompt, and afterglow phase. Since the afterglow phase is only rarely observed by \textit{Fermi}-GBM, this phase is not directly used in the analysis. The duration of the total signal region is called $T$. It is defined by the time interval from the start of the total signal region (start of precursor or prompt phase) until the end of the signal region (end of prompt phase or afterglow). The distribution of $T$ is shown in Fig. \ref{fig:grb_distribution_T}. When we investigate only the precursor or prompt phase, $T$ is limited to the duration of the considered phase region. The details of this procedure are outlined in Appendix \ref{app:phaseidentification}. An illustration is shown in Fig. \ref{fig:lightcurve_example}. For ten GRBs, our procedure fails to detect a photon-count excess above background. These GRBs cannot be analyzed and are removed from the sample, as mentioned in Sec. \ref{subsec:dataselection}. To include only the regions of interest, we apply a window function to the light curve. This window isolates the identified signal regions and is defined as

    \begin{equation}
        w(n) = \begin{cases}
        \frac{1}{2} ( 1 - \cos (\frac{2\pi n}{\alpha N})) & 0 \leq n \leq \frac{\alpha N}{2}, \cr
        1 & \frac{\alpha N}{2} \leq n \leq \frac{N}{2}, \cr
        w(N-n) & \frac{N}{2} \leq n \leq N,
        \end{cases}
    \end{equation}

    often referred to as the cosine-tapered window or \textit{Tukey window} \cite{Harris_1978}. Here, $N$ is the window length, $\alpha$ the cosine fraction, and $n$ a data point in [$0$,$N$]. The parameter $\alpha$ reflects the fraction of the window that is contained in the cosine-tapered region. A Tukey window with $\alpha \xrightarrow{} 0$ represents a rectangular function, while $\alpha \xrightarrow{} 1$ leads to a Hann window \cite{Harris_1978}. We choose $\alpha = 0.25$. We want to ensure that no power of the signal regions is lost in the cosine tapering. Therefore, we apply the window on the signal region, extended by 12.5\% on the left and right. Subsequently, the trace is extended with zeros on both sides (i.e.,\ zero padding) to efficiently perform the fast Fourier transform (FFT). When analyzing a group of GRBs, we take the closest power of two for the longest burst in the group for the length of the zero-padded trace. 

    The amplitudes $F_k$ and $P_k$ of the discrete Fourier transform (DFT) and power-density spectrum, generated from the zero-padded light curve $x_n$ with length $N$, are calculated using the Leahy normalization \cite{Leahy_1983}: 

    \begin{align}
        F_k &= \sum^{N-1}_{n=0} x_n \cdot e^{-2\pi i \frac{k}{N} n} \label{eq:fouriertransform}\\
        P_k &= \frac{2}{N_{ph}} \lvert F_k \rvert ^2, \label{eq:leahynormalisation}
    \end{align}

    \noindent where $N_{ph}$ is the sum of the photon counts. Following the methodology outlined in Ref. \cite{Guidorzi_2012}, we subtract the statistical noise from the individual PDS. The noise is estimated at high frequencies (50 Hz $< f <$ 100 Hz) by fitting a straight line to the data in this range. We calculate the mean value of the noise at $f = 50$ Hz and $f = 100$ Hz and subtract this mean value from the PDS across all frequency bins. Each PDS is then normalized to the peak of its corresponding light curve, ensuring that all PDSs contribute equally to the group's average. The Leahy normalization, as defined in Eq. \eqref{eq:leahynormalisation}, enables us to determine the uncertainties on the individual PDS, following the approach described in Ref. \cite{Guidorzi_2011}: 

    \begin{equation}
    \label{eq:individualerror}
        \sigma_{P_j} = 
        \begin{cases}
            2 \cdot \sqrt{P_j + 1} & j < N/2, \\
            2 \cdot \sqrt{2(P_j + 1)} & j = N/2.
        \end{cases}
    \end{equation}

    The average PDS will be calculated for a certain population of GRBs, as will be explained in Sec. \ref{sec:results}. The averaging of different individual power-density spectra is not trivial. It is based on the strong assumption that the different time spectra of GRBs are the realizations of one common stochastic process. In other words, one unique process is assumed to give rise to the different temporal characteristics that GRBs display. The average power density is calculated per frequency bin, and its uncertainties are the corresponding standard deviations, i.e.,

    \begin{align}
        \Bar{P} \, [f] &= \sum_{j} \, \frac{P_j \, [f]}{N_{G}} \ , \\
        \sigma_{\Bar{P}} \, [f] &= \frac{1}{\sqrt{N_{G}}} \,\sqrt{\sum_{j} \, \frac{(P_j[f] - \Bar{P} [f]) ^2}{N_G-1}} \ , \label{eq:erroronthemean}
    \end{align}

    where the summation is performed over the individual power-density spectra and $N_G$ is the total number of GRBs in the sample \cite{Lista_2023}. The final uncertainty on the average PDS is the combination of the error of the individual spectra [Eq. \eqref{eq:individualerror}] and the error on the mean [Eq. \eqref{eq:erroronthemean}].

\subsection{\label{subsec:pdsmodelling} PDS modeling}

    The average PDS is modeled by a smooth broken power-law fit, as proposed in \cite{Guidorzi_2012}:

    \begin{equation}
    \label{eq:smoothbrokenpowerlaw}
        \mathrm{P} \, [f] = 2^{1/n} \cdot F_0 \cdot \left[ \left( \frac{f}{f_b} \right)^{n \beta_{LF}} + \left( \frac{f}{f_b} \right)^{n \beta_{HF}} \right]^{-1/n}.
    \end{equation}

    Here, the function is given in terms of the break frequency $f_b$, the amplitude at the break frequency $F_0$, the low- and high-frequency power-law indices, respectively, $\beta_{LF}$ and $\beta_{HF}$, and the peakedness parameter $n$, which is the only parameter that is fixed. We take $n = 10$, implying a rather sharp break around $f_b$, to compare with previous research in Refs. \cite{Guidorzi_2012, Dichiara_2013a, Zhou_2024}. As described in Ref. \cite{Guidorzi_2012}, a varying value of $n$ does not impact the results. 
    
    In general, Eq. \eqref{eq:smoothbrokenpowerlaw} is fitted in the frequency interval $f = [0.01; 1.00]$ Hz. This is the frequency interval in which we expect most of the power-law features to appear. Typically, the frequency corresponding to the longest burst appears at $f < 0.01$ Hz. For example, the longest burst in the sample of 159 LGRBs corresponds to a frequency of $f = 0.003$ Hz. However, at $f < 0.01$ Hz, the average PDS flattens quickly, and the spectrum cannot be characterized by (part of) a broken power law. Since we are more interested in the behavior of the high-frequency slope, we decide to set $f = 0.01$ Hz as a lower-frequency bound. Note, however, that for some GRB groups the duration of the longest burst can be $< 100$ s. In that case, we take the corresponding frequency, $f = 1/T$, as low-frequency bound. The high-frequency bound is set at $f = 1 \ \mathrm{Hz}$ to avoid the noise-dominated region. Previous studies have identified a second break in the average PDS of long bursts around $1$–$2$ Hz \cite{Beloborodov_2000, Dichiara_2013a}. However, since our fitting interval is limited to frequencies below $1$ Hz, this analysis is not sensitive to the presence of a break in that region.  
    
    As can be seen from the spectrum for the long bursts in Fig. \ref{fig:PDS_allGRBs}, the general shape of the PDS is dominated at low frequencies by a sinc function, given by $\mathrm{sinc}(x) = \sin(x)/x$. This is expected, since the light curve can be approximated by a top hat function in temporal space. The saturation frequency at low frequencies corresponds to the length of the zero-padded time trace. When $f \gtrsim 1$ Hz, the noise starts distorting the shape. In between, we expect the features due to the time variability of the burst. Equation \eqref{eq:smoothbrokenpowerlaw} is fitted to the data using a standard least-squares formalism. The model's goodness of fit is then parametrized by the reduced chi-squared statistic $\chi^2$/$n_{d.o.f.}$, where we divided the Pearson's chi-squared test statistic by the number of degrees of freedom $n_{d.o.f.}$. The least-squares formalism provides the best-fit parameters together with the statistical errors on the parameters. In this paper, we will refer to those errors as the statistical errors of the fit. This uncertainty is reported as the first error term on the fitting parameters. 
    
    As will become clear in the following sections, the power-law index can be very dependent on the upper limit of the fitting interval. To account for the dependence on the chosen upper limit, we add a systematic error to the best-fitting parameters. The systematic error is based on the spread of the best-fitting parameters between $f = 0.8$ Hz and $f = 1.2$ Hz. This uncertainty is reported as the second error term on the fitting parameters.

\section{\label{sec:results} RESULTS}


   \begin{sidewaystable*}
        \centering
        \caption{Best-fit parameters $\chi^2/n_{d.o.f.}$, $\beta_{LF}$, $\beta_{HF}$, $f_b$, $F_0$ of the average PDS for different groups of GRBs (with $N$ GRBs). If not mentioned differently, the spectra are fitted with a broken power law (BPL). If the spectrum is fitted with a single power law (SPL), the exponent enters in the $\beta_{HF}$ column, and the multiplication factor in the $F_0$ column. Abbreviations: SPL, single power law; OF, observer frame; RF, redshift-corrected frame.}
        \begin{tabular}{p{4.7cm} p{0.75cm} p{1.3cm} p{4.0cm} p{4.0cm} p{4.0cm} p{4.0cm} }
        \toprule
           GRB set & $N$ & $\chi^2/n_{d.o.f.}$ & $\beta_{LF}$ & $\beta_{HF}$ & $f_{b}$ [Hz] & $F_{0} \ [\mathrm{Hz^2}]$ \\
           \midrule 
           \multicolumn{7}{l}{\cellcolor[HTML]{ffd1b2}{\color[HTML]{000000} \textbf{Short vs. long GRBs}}} \\
                Long GRBs            & 159 & 0.42 & $-0.95 \pm 0.02 \pm 0.02$ & $-1.90 \pm 0.01 \pm 0.07$    & $0.098 \pm 0.004 \pm 0.009$ & $107 \pm 6 \pm 12$ \\ 
                Short GRBs (RF, SPL) & 24  & 0.13 & ...                         & $-1.337 \pm 0.007 \pm 0.083$ & ...                           & $33.4 \pm 0.4 \pm 2.8$ \\  
                Short GRBs (OF, BPL) & 395 & 0.27 & $-0.87 \pm 0.02 \pm 0.02$ & $-1.71 \pm 0.03 \pm 0.09$    & $3.2 \pm 0.2 \pm 0.4$       & $5.8 \pm 0.4 \pm 0.8$ \\ 
                
           \midrule
           \multicolumn{7}{l}{\cellcolor[HTML]{ffd1b2}{\color[HTML]{000000} \textbf{Evolution with duration}}} \\
                2 s  $< T_{90} <$ 22 s  & 51  & 0.37 & $-0.55 \pm 0.02 \pm 0.03$ & $-1.92 \pm 0.02 \pm 0.09$ & $0.109 \pm 0.003 \pm 0.009$ & $171 \pm 7 \pm 15$ \\
                22 s $< T_{90} <$ 52 s  & 48  & 0.26 & $-1.07 \pm 0.04 \pm 0.04$ & $-1.94 \pm 0.02 \pm 0.03$ & $0.063 \pm 0.005 \pm 0.006$ & $164 \pm 20 \pm 26$ \\
                52 s $< T_{90} <$ 200 s & 51  & 0.30 & $-1.22 \pm 0.06 \pm 0.02$ & $-1.87 \pm 0.03 \pm 0.04$ & $0.069 \pm 0.009 \pm 0.007$ & $96 \pm 21 \pm 16$  \\
           \midrule
                2 s  $< T <$ 12 s (BPL)  & 50  & 0.31 & $-1.24 \pm 0.08 \pm 0.12$ & $-2.01 \pm 0.03 \pm 0.14$ & $0.16 \pm 0.01 \pm 0.03$ & $82 \pm 12 \pm 34$ \\
                12 s $< T <$ 27 s (BPL)  & 51  & 0.20 & $-1.53 \pm 0.04 \pm 0.21$ & $-2.31 \pm 0.07 \pm 0.42$ & $0.25 \pm 0.03 \pm 0.18$ & $14 \pm 3 \pm 38$ \\
                27 s $< T <$ 120 s (BPL) & 50  & 0.34 & $-1.17 \pm 0.06 \pm 0.02$ & $-1.73 \pm 0.03 \pm 0.05$ & $0.08 \pm 0.01 \pm 0.01$ & $94 \pm 22 \pm 16$  \\        
           \midrule
                2 s  $< T <$ 12 s (SPL)  & 50  & 0.44 & ... & $- 1.91 \pm 0.02 \pm 0.09$ & ... & $2.35 \pm 0.05 \pm 0.38$ \\
                12 s $< T <$ 27 s (SPL)  & 51  & 0.21 & ... & $- 2.06 \pm 0.03 \pm 0.02$ & ... & $0.73 \pm 0.03 \pm 0.02$ \\
                27 s $< T <$ 120 s (SPL) & 50  & 0.37 & ... & $- 1.73 \pm 0.03 \pm 0.05$ & ... & $1.15 \pm 0.04 \pm 0.09$  \\      
           \midrule 
           \multicolumn{7}{l}{\cellcolor[HTML]{ffd1b2}{\color[HTML]{000000} \textbf{Evolution with peak rate}}} \\
                $0 < \mathrm{PR} < 3.5 \cdot 10^4$ c/s             & 49 & 0.11 & $-0.997 \pm 0.031 \pm 0.006$ & $-2.18 \pm 0.03 \pm 0.05$ & $0.122 \pm 0.007 \pm 0.005$ & $70 \pm 6 \pm 3$  \\
                $3.5 \cdot 10^4 < \mathrm{PR} < 15 \cdot 10^4$ c/s & 53 & 0.18 & $-1.021 \pm 0.038 \pm 0.002$ & $-1.78 \pm 0.03 \pm 0.01$ & $0.091 \pm 0.008 \pm 0.002$ & $105 \pm 14 \pm 3$ \\
                $15 \cdot 10^4 < \mathrm{PR} < 150 \cdot 10^4$ c/s & 50 & 0.56 & $-0.79 \pm 0.05 \pm 0.05$    & $-1.86 \pm 0.02 \pm 0.09$ & $0.078 \pm 0.004 \pm 0.010$ & $178 \pm 13 \pm 30$  \\

           \midrule 
           \multicolumn{7}{l}{\cellcolor[HTML]{ffd1b2}{\color[HTML]{000000} \textbf{Evolution with fluence}}} \\
                $0 < \mathrm{F} < 7 \cdot 10^{-6}$ erg/$\mathrm{cm^2}$ & 53 & 0.19 & $-0.89 \pm 0.03 \pm 0.03$ & $-1.92 \pm 0.03 \pm 0.16$ & $0.116 \pm 0.006 \pm 0.021$ & $74 \pm 6 \pm 17$  \\
                $7 \cdot 10^{-6} < \mathrm{F} < 25 \cdot 10^{-6}$ erg/$\mathrm{cm^2}$ & 52 & 0.18 & $-0.971 \pm 0.030 \pm 0.006$ & $-2.00 \pm 0.03 \pm 0.04$ & $0.101 \pm 0.006 \pm 0.005$ & $92 \pm 8 \pm 5$  \\
                $25 \cdot 10^{-6} < \mathrm{F} < 400 \cdot 10^{-6}$ erg/$\mathrm{cm^2}$ & 48 & 0.19 & $-0.95 \pm 0.03 \pm 0.03$ & $-1.88 \pm 0.02 \pm 0.10$ & $0.095 \pm 0.006 \pm 0.014$ & $139 \pm 13 \pm 27$  \\
           
           \midrule 
           \multicolumn{7}{l}{\cellcolor[HTML]{ffd1b2}{\color[HTML]{000000} \textbf{Evolution with redshift}}} \\
                0.00 $<$ z $<$ 0.90 & 53 & 0.24 & $-1.04 \pm 0.04 \pm 0.01$ & $-1.88 \pm 0.03 \pm 0.05$ & $0.112 \pm 0.009 \pm 0.008$ & $77 \pm 10 \pm 7$ \\
                0.90 $<$ z $<$ 1.80 & 54 & 0.17 & $-0.94 \pm 0.05 \pm 0.13$ & $-1.68 \pm 0.02 \pm 0.13$ & $0.063 \pm 0.006 \pm 0.026$ & $166 \pm 23 \pm 96$  \\
                1.80 $<$ z $<$ 5.00 & 51 & 0.24 & $-0.82 \pm 0.03 \pm 0.02$ & $-2.20 \pm 0.02 \pm 0.07$ & $0.111 \pm 0.004 \pm 0.007$ & $119 \pm 7 \pm 10$  \\
           \midrule
           
           \multicolumn{7}{l}{\cellcolor[HTML]{ffd1b2}{\color[HTML]{000000} \textbf{Different phases}}} \\
                Precursors (BPL) & 26  & 0.06 & $-1.1726 \pm 0.0428 \pm 0.0003$ & $-1.72 \pm 0.24 \pm 0.03$    & $1.71 \pm 0.64 \pm 0.03$    & $3.08 \pm 1.57 \pm 0.04$  \\
                Precursors (SPL) & 26  & 0.11 & ...                               & $-1.262 \pm 0.023 \pm 0.005$ & ...                           & $5.38 \pm 0.18 \pm 0.04$  \\
                Prompt phase     & 159 & 0.32 & $-0.93 \pm 0.02 \pm 0.01$       & $-1.76 \pm 0.01 \pm 0.04$    & $0.083 \pm 0.003 \pm 0.006$ & $139 \pm 8 \pm 12$ \\
           
            \bottomrule 
        \end{tabular}
        \label{tab:fittingparameters}
    \end{sidewaystable*}

\subsection{\label{subsec:allGRBs} Full GRB sample}

    We aim to analyze and compare the characteristics of the average PDS, with particular focus on the high-frequency slope, across different types of GRBs. The analysis distinguishes GRBs based on duration, photon peak rate, and observed fluence (as proxies for brightness), redshift, and the different phases of their light curves (precursor, prompt, and noise emission). As a reference, we compute the average PDS of the full GRB sample, including both short and long GRBs. Since these two classes are thought to originate from different progenitors, a key question arises: Can they be analyzed together under the assumption that their gamma-ray emission is produced by similar underlying mechanisms? 
    
\begin{figure}
        \centering
        \includegraphics[width=0.95\linewidth]{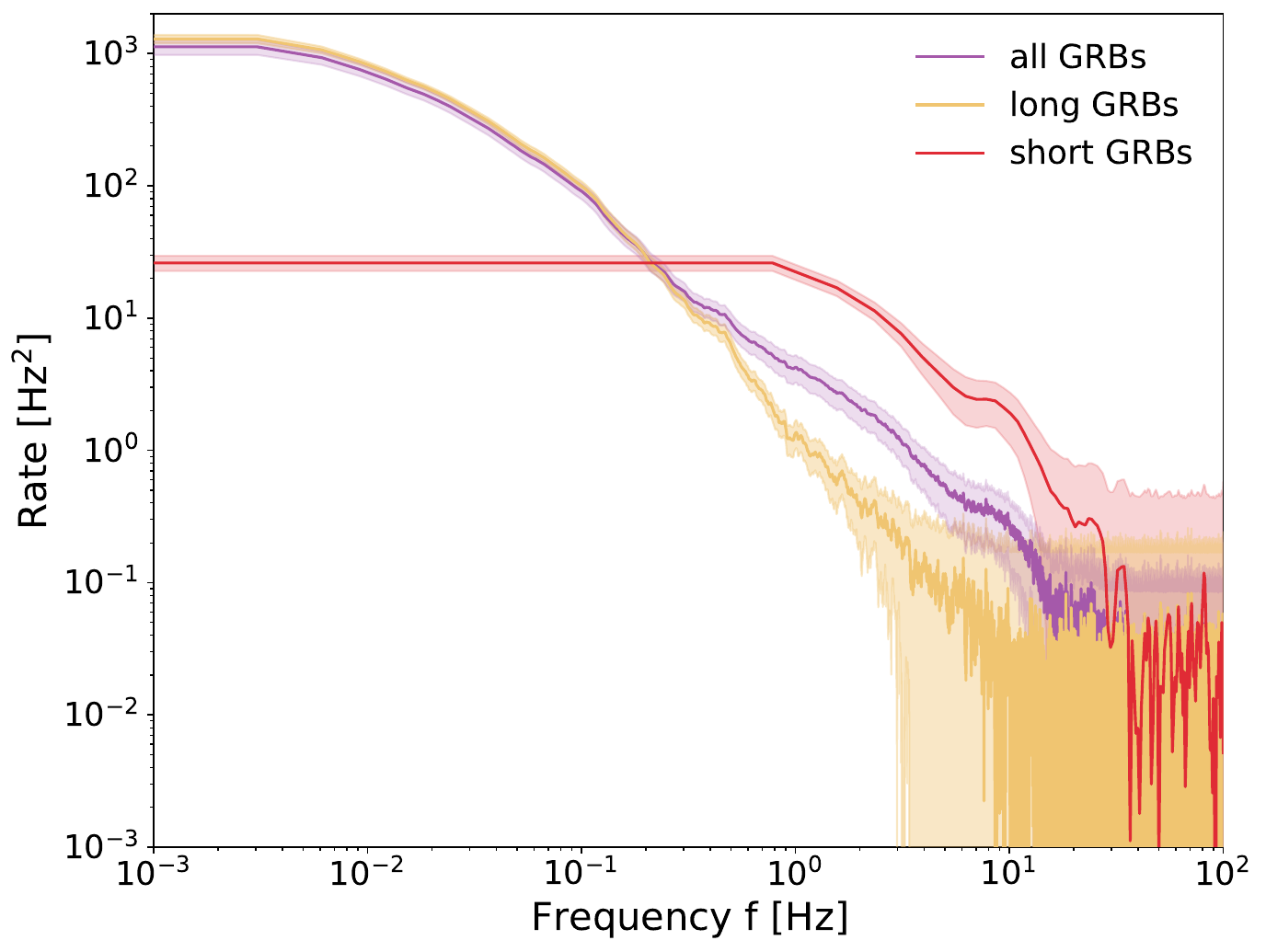}
        \caption{The average PDS of long and short bursts (purple) is a combination of the characteristics of the PDS of short GRBs (red) and long GRBs (yellow). As both are significantly different, they are analyzed separately.}
        \label{fig:PDS_allGRBs}
    \end{figure}

    We examine the contribution of short and long GRBs to the combined average PDS in Fig. \ref{fig:PDS_allGRBs}, following the method described in Sec. \ref{subsec:dataprocessing}. They are classified as short or long based on their $T_{90}$: $T_{90} > 2$ s for LGRBs and $T_{90} < 2$ s for SGRBs \cite{Kouveliotou_1993}. Additionally, we require that each GRB is correctly identified by our phase-identification procedure (outlined in Appendix \ref{app:phaseidentification}) as long or short, excluding, e.g., possible long extended emission of short bursts. Under this criterion, $18$ long GRBs and $3$ short GRBs are excluded from the sample. Finally, our sample comprises 24 SGRBs and 159 LGRBs, resulting in a combined average PDS (long + short) of 183 GRBs. Notably, the short bursts start to contribute only at frequencies above $f \sim 0.3$ Hz and significantly distort the spectrum at higher frequencies. Analyzing the power-law index of the combined average PDS could lead to either an underestimation or overestimation of the true power-law index. Therefore, we decide to analyze the PDS of short and long GRBs separately. 

\subsection{\label{subsec:longGRBs} Long GRBs}

\begin{figure}
    \centering
        \includegraphics[width=0.95\linewidth]{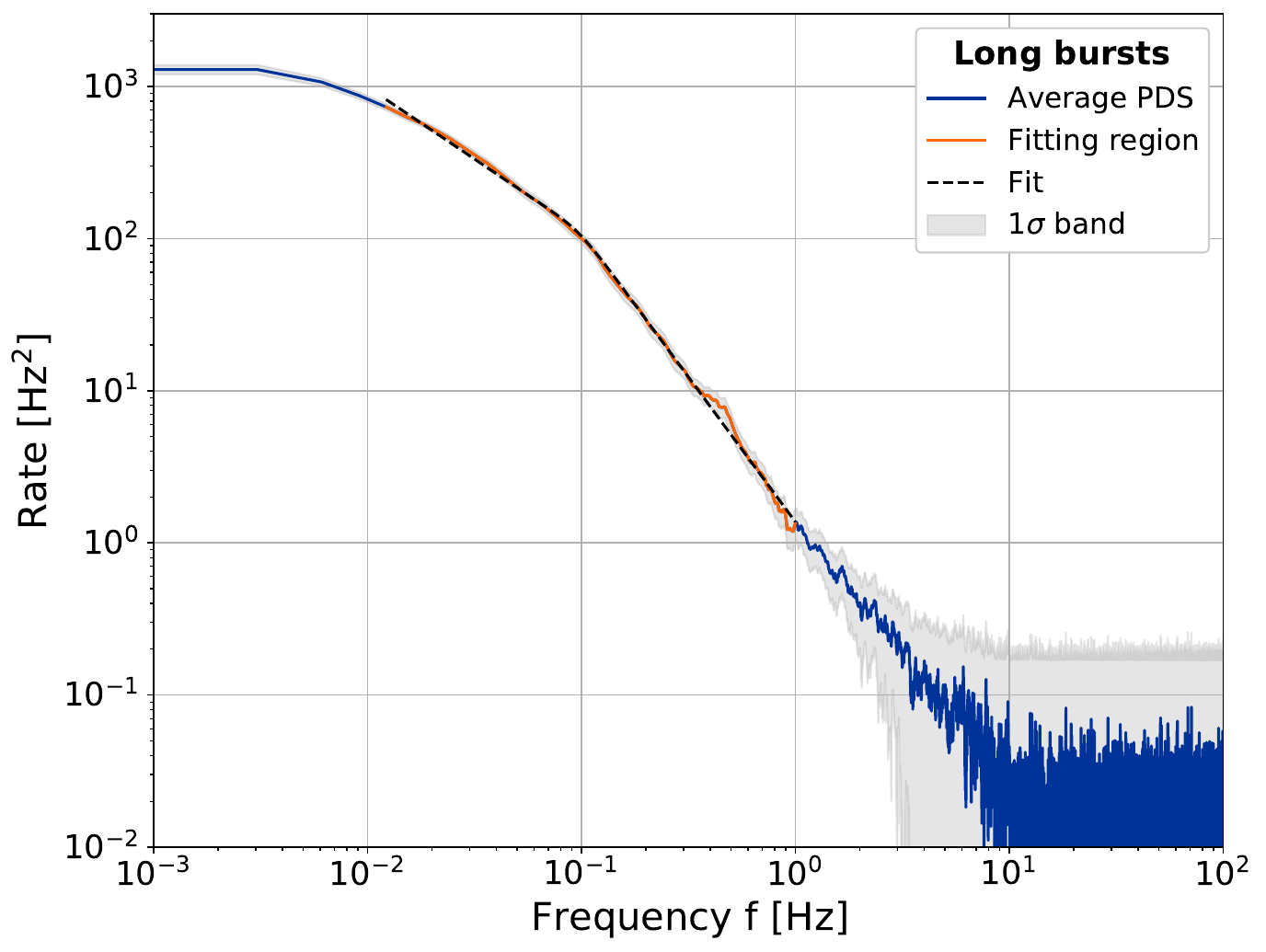}
        \caption{Average PDS for 159 LGRBs (blue). The gray band denotes the $1\sigma$-uncertainty interval on the PDS. The best fit in the orange region is given in black.}
        \label{fig:PDS_L0}
    \end{figure}

    The population of the LGRBs is selected as explained above, and contains 159 GRBs. The spectrum is shown in Fig. \ref{fig:PDS_L0}. The broken power-law model of Eq. \eqref{eq:smoothbrokenpowerlaw} is fitted between 0.01 and 1 Hz. The best-fit parameters are listed in Table \ref{tab:fittingparameters}. The low- and high-frequency power-law indices become $\beta_{LF} = - 0.95 \pm 0.02 \pm 0.02$ and $\beta_{HF} = - 1.90 \pm 0.01 \pm 0.07$. The break in the spectrum occurs at $f_b = 0.098 \pm 0.004 \pm 0.009$ Hz.

    The durations of the longest burst and shortest burst within this group are, respectively, $T_{LB} = 279$ s and $T_{SB} = 2.76$ s, corresponding to a frequency of $f_{LB} = 1/T_{LB} = 0.0036$ Hz and $f_{SB} = 1/T_{SB} = 0.36$ Hz. The median duration is $T_{MD} = 23$ s, or $f_{MB} = 1/T_{MD} = 0.044$ Hz. Notice that the power-law features extend to higher frequencies, reflecting the internal structures of the time patterns, still exposing a power law. 

\subsubsection{\label{subsubsec:peakrate} Effect of GRB fluence and peak rate}

    The brightness of a GRB can be characterized by two parameters: the observed fluence and the peak photon rate of the light curve. Fluence represents the total energy received per unit surface. It is provided by the GBM instrument and is based on the 10-1000 keV energy band for each burst. In contrast, the peak photon rate is derived from our phase-identification procedure (Appendix \ref{app:phaseidentification}). It is defined as the rate of the peak of the redshift-corrected Bayesian block histogram constructed during phase identification. While the fluence can be high for long but dim GRBs, the peak photon rate reflects brief, intense brightening events. When investigating the correlation between observed fluence $\mathrm{F}$ and peak photon rate $\mathrm{PR}$ we see that the two parameters are not directly proportional. The Pearson's correlation coefficient \cite{Pearson_1896}, Spearman's rank correlation coefficient \cite{Spearman_1904}, and Kendall's rank correlation \cite{Kendall_1990} coefficient all indicate a weak positive correlation ($\rho = 0.05$, $r = 0.14$, and $\tau = 0.10$, respectively). Therefore, observed fluence and peak rate are investigated independently. 

\begin{figure}
        \centering
        \includegraphics[width=0.95\linewidth]{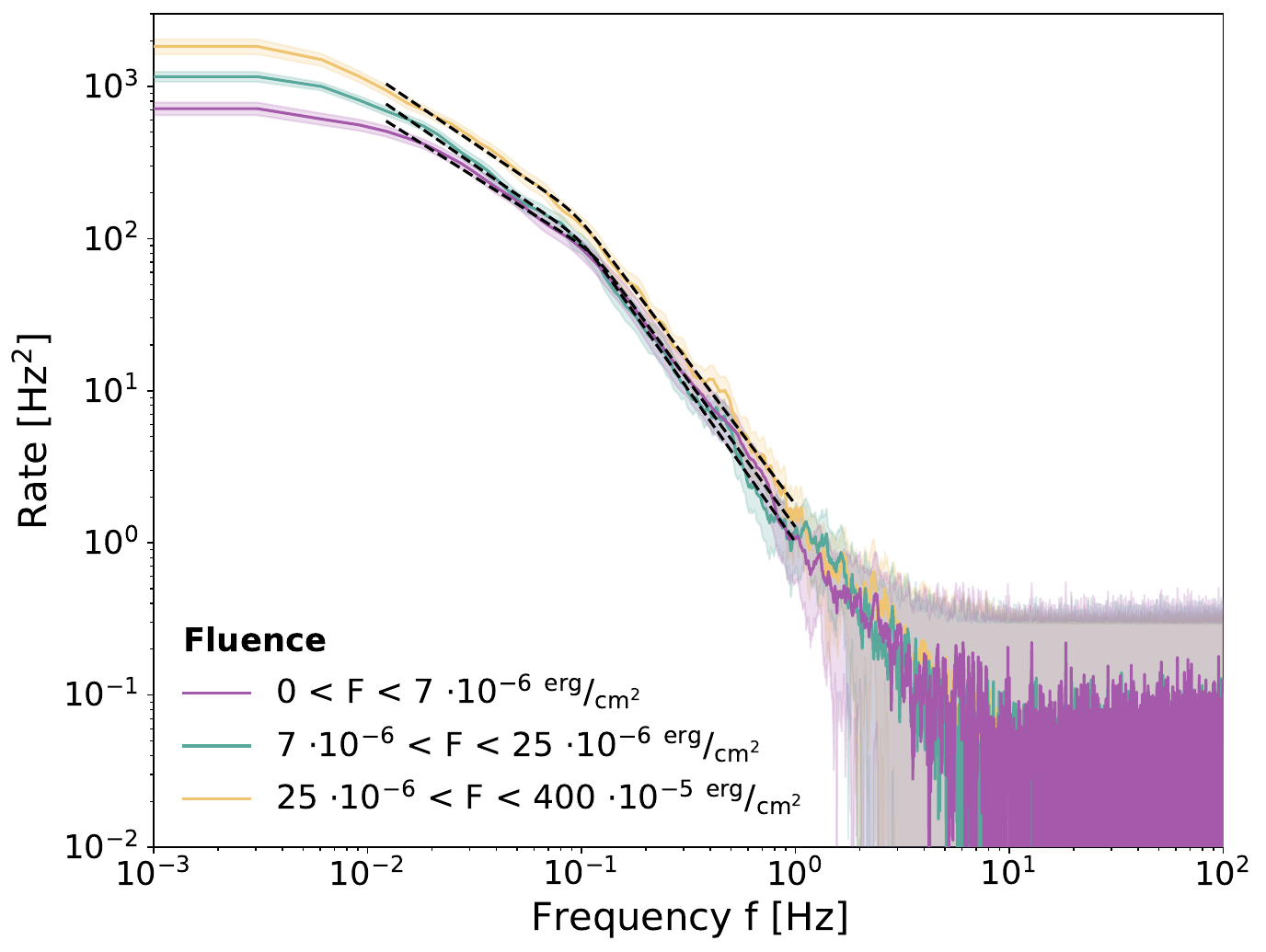}
        \caption{Average PDS for three fluence groups of LGRBs. The band around the PDS gives the $1\sigma$-error band. The dotted lines represent the best power-law fits.}
        \label{fig:combined_fluence}
    \end{figure}

    Based on the fluence distribution shown in Fig. \ref{fig:grbdistribution}, we exclude six outliers with $\mathrm{F} > 4.0 \cdot 10^{-4}$ erg/$\mathrm{cm^2}$, leaving a total of 153 long GRBs. The sample is subsequently divided into three fluence groups, which are chosen such that the number of GRBs in each group approximates 50 GRBs: $0 < \mathrm{F} < 7 \cdot 10^{-6}$ erg/$\mathrm{cm^2}$ (53 LGRBs), $7 \cdot 10^{-6} < \mathrm{F} < 25 \cdot 10^{-6}$ erg/$\mathrm{cm^2}$ (52 LGRBs), and $25 \cdot 10^{-6} < \mathrm{F} < 400 \cdot 10^{-6}$ erg/$\mathrm{cm^2}$ (48 LGRBs). The average power-density spectra for these groups are shown in Fig. \ref{fig:combined_fluence}. The corresponding best-fit parameters are listed in Table \ref{tab:fittingparameters}. All spectra show similar features. The total amplitude of the average PDS increases with fluence, because higher fluence GRBs on average have light curves with more photons. The high-frequency power-law indices, ordered from low to high fluence, are $\beta_{HF} = - 1.92 \pm 0.03 \pm 0.16$, $\beta_{HF} = - 2.00 \pm 0.03 \pm 0.04$, and $\beta_{HF} = - 1.88 \pm 0.02 \pm 0.10$. These values are consistent with each other.

\begin{figure}
        \centering
        \includegraphics[width=0.95\linewidth]{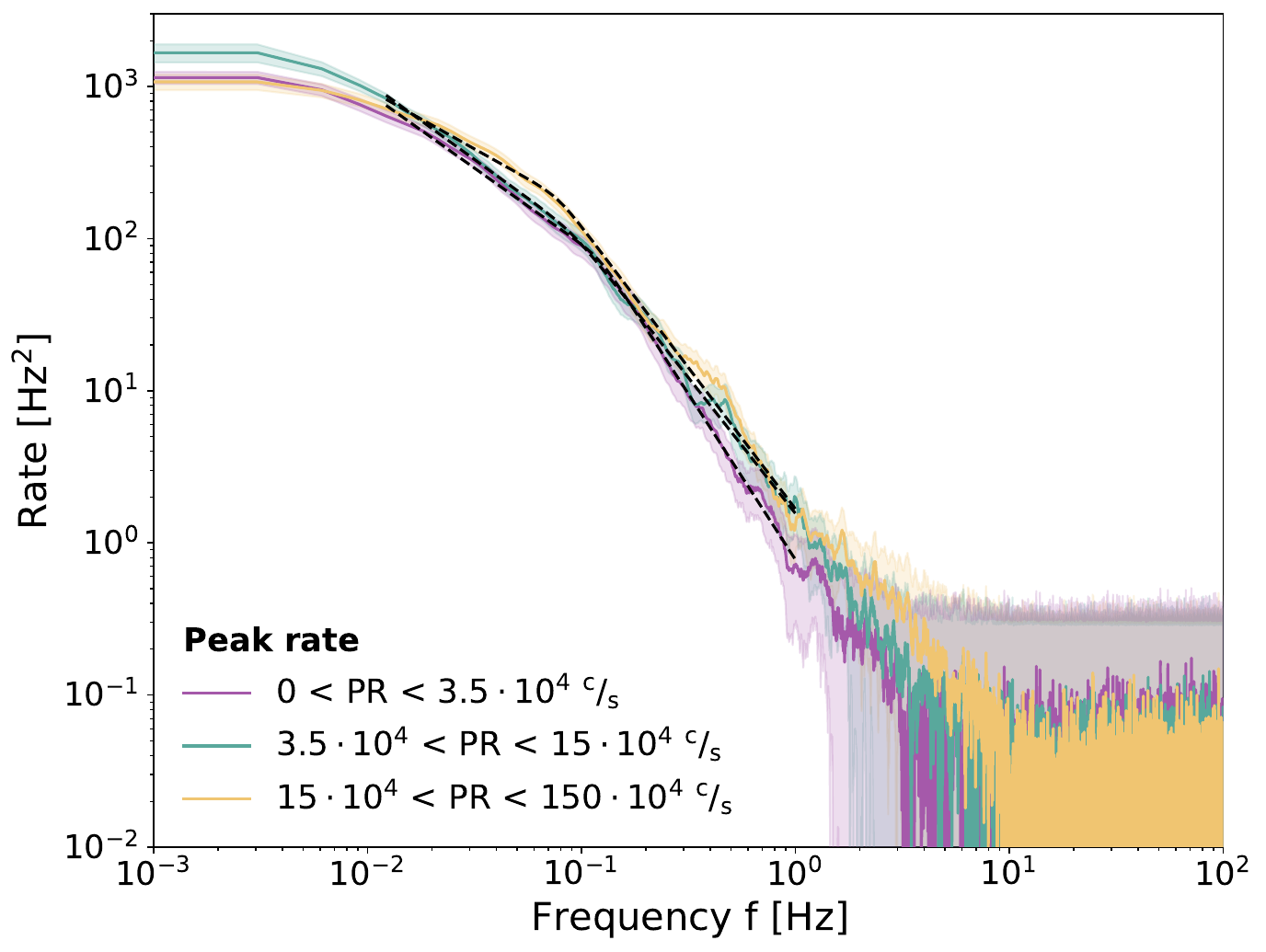}
        \caption{Average PDS for three peak-rate groups of LGRBs. The band around the PDS gives the $1\sigma$-error band. The dotted lines represent the best power-law fits.}
        \label{fig:combined_peakrate}
    \end{figure}

    Similarly, the long GRBs are grouped based on their peak rate. To minimize large differences within a single bin, we limit the peak rate to  $\mathrm{PR} = 150 \cdot 10^4$ counts/s. This leads to the exclusion of seven GRBs and leaves a total of 152 long GRBs for the analysis. We identify the following three groups of approximately 50 GRBs: $0 < \mathrm{PR} < 3.5 \cdot 10^4$ c/s (49 LGRBs), $3.5 \cdot 10^4 < \mathrm{PR} < 15 \cdot 10^4$ c/s (53 LGRBs), and $15 \cdot 10^4 < \mathrm{PR} < 150 \cdot 10^4$ c/s (50 LGRBs). The corresponding spectra are shown in Fig. \ref{fig:combined_peakrate}, with the best-fit results listed in Table \ref{tab:fittingparameters}. The high-frequency power-law indices, ordered from low to high peak rate, are $\beta_{HF} = - 2.18 \pm 0.03 \pm 0.05$, $\beta_{HF} = - 1.78 \pm 0.03 \pm 0.01$, and $\beta_{HF} = - 1.86 \pm 0.02 \pm 0.09$. Notably, the group with the lowest peak rate exhibits a steeper spectrum compared to the others. 
    
\subsubsection{\label{subsubsec:duration} Effect of GRB duration}

\begin{figure}
        \centering
        \includegraphics[width=0.95\linewidth]{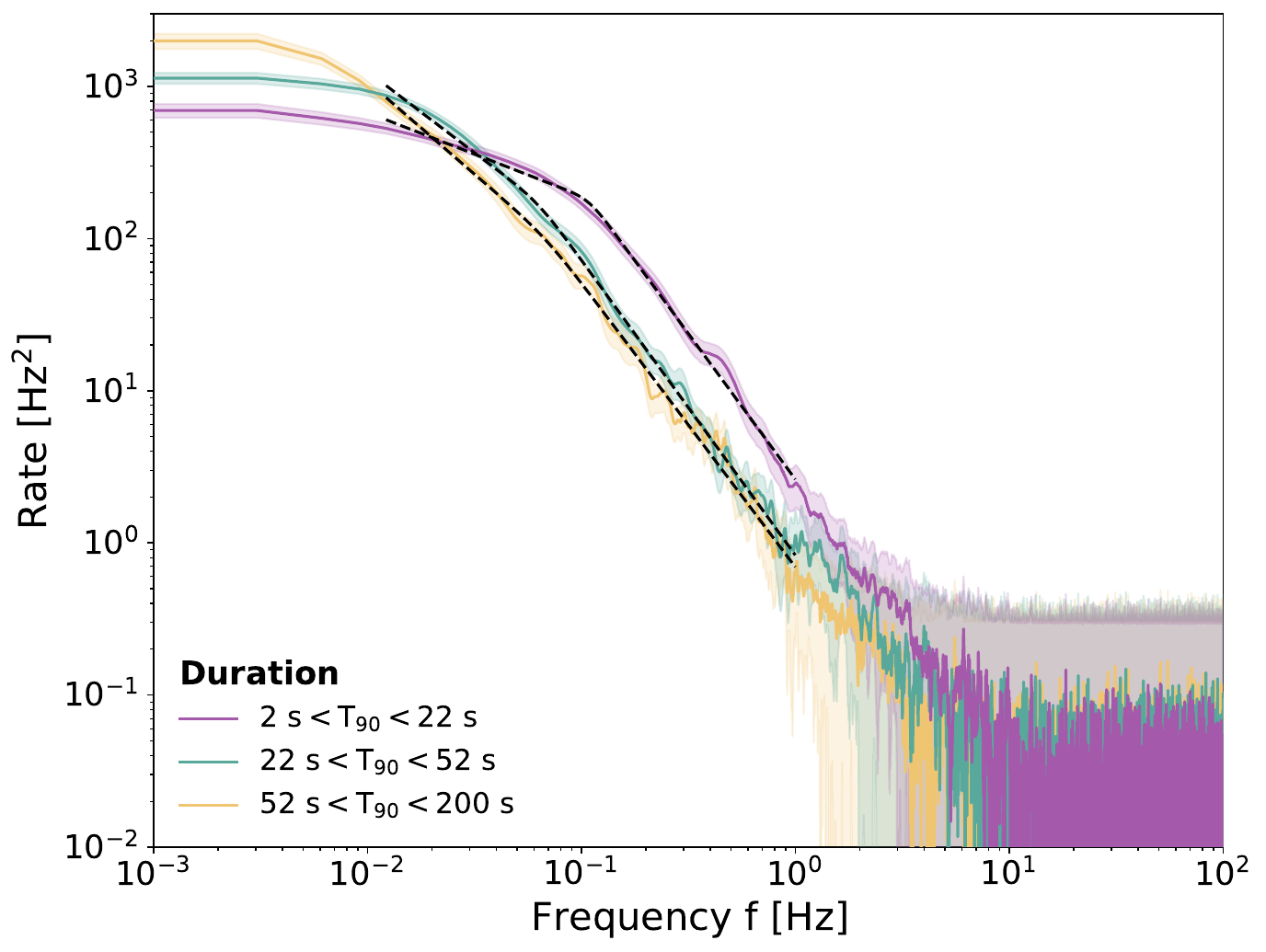}
        \caption{Average PDS for three duration groups of LGRBs, based on their $T_{90}$. The band around the PDS gives the $1\sigma$-error band. The dotted lines represent the best power-law fits.}
        \label{fig:combined_duration}
    \end{figure}

    At the beginning of this section, we already distinguished between short and long bursts, based on the $T_{90}$ and our phase identification. Now, we further divide the long GRBs into three groups of around 50 GRBs, based on their observed $T_{90}$. Nine GRBs have $T_{90} > 200$ s and are regarded as outliers, based on the distribution in Fig. \ref{fig:grbdistribution}. They are excluded from the sample. The following subgroups are defined: 2.0 s  $< T_{90} <$ 22 s (51 LGRBs), 22.0 s $< T_{90} <$ 52 s (48 LGRBs), and 52.0 s $< T_{90} <$ 200 s (51 LGRBs). The best-fit results can be found in Table \ref{tab:fittingparameters} and the spectra in Fig. \ref{fig:combined_duration}. Notice that the PDS has the largest amplitude for the longest bursts and receives more power at the lowest frequencies. This influences the low-frequency power-law index, but our results show that the high-frequency power-law index remains constant within uncertainties over the different $T_{90}$ bins: $\beta_{HL} = -1.92 \pm 0.02 \pm 0.09$, $\beta_{HL} = -1.94 \pm 0.02 \pm 0.03$, and $\beta_{HL} = -1.87 \pm 0.03 \pm 0.04$ (for increasing mean $T_{90}$). 

    However, in this analysis, the main temporal characteristic of the GRB light curve is the duration of the redshift-corrected signal region, $T$, or the duration of the cosine-tapered window. This differs from $T_{90,c}$, where the subscript ``c" denotes the redshift-corrected $T_{90}$. The $T_{90}$ for the \textit{Fermi}-GBM instrument is computed in the 50-300 keV energy range \cite{von_Kienlin_2020}, while our analysis uses light curves in the 8 keV to 1 MeV energy band. Still, correlation coefficients indicate a correlation between the duration of the signal zone $T$ and $T_{90,c}$: $\rho = 0.50$, $r = 0.66$, and $\tau = 0.52$. This correlation, however, weakens with $\sim 10$\% for observed $T_{90}$. As the GRBs were classified based on observed $T_{90}$ earlier, we study the dependence of the PDS slopes on $T$ as well. 
    
\begin{figure}
        \centering
        \includegraphics[width=0.95\linewidth]{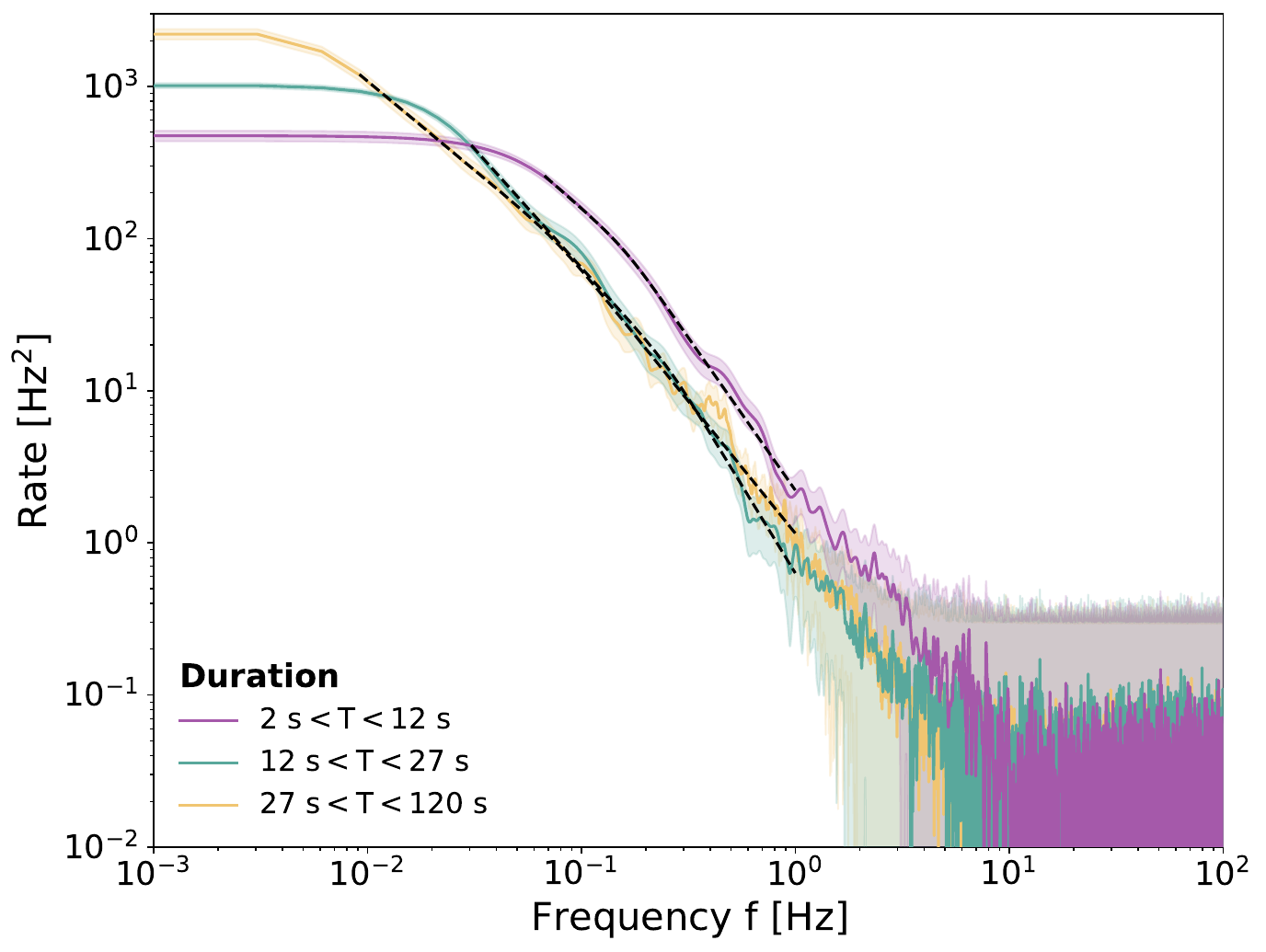}
        \caption{Average PDS for three duration groups of LGRBs, based on their $T$. The band around the PDS gives the $1\sigma$-error band. The dotted lines represent the best power-law fits.}
        \label{fig:combined_durationT}
    \end{figure}

    We exclude nine outliers with $T > 120$ s based on the distribution shown in Fig. \ref{fig:grb_distribution_T}, leaving a total of 151 long GRBs. We divide them in three groups of approximately 50 GRBs: 2 s  $< T <$ 12 s (50 LGRBs), 12 s $< T <$ 27 s (51 LGRBs), and 27 s $< T <$ 120 s (50 LGRBs). Their spectra are shown in Fig. \ref{fig:combined_durationT} and the corresponding best-fit results are listed in Table \ref{tab:fittingparameters}. The duration of the longest bursts are, respectively, $T_{LB} = 15.47$ s, $T_{LB} = 35.85$ s, and $T_{LB} = 138.45$ s. That means that the start of the fitting interval moves to lower frequencies for longer $T$. Indeed, the relation between the duration and the saturation frequency appears more distinct now. The longest bursts have again the largest amplitude in the power spectra. The high-frequency power-law indices are, for increasing $T$, $\beta_{HL} = -2.01 \pm 0.03 \pm 0.14$, $\beta_{HL} = -2.31 \pm 0.07 \pm 0.42$, and $\beta_{HL} = -1.74 \pm 0.03 \pm 0.05$. Note that large differences exist between the absolute values of the slopes, including large errors. For example, the systematic error of the mid-long group is large because of the large differences in the power-law indices when fitting between $f = [0.8; 1.2]$ Hz. In addition, the low-frequency slope is small, compared to other subsets. Visually, one can see that the two shortest-duration groups lack a clear break in the spectrum. Taking all those features together, we have decided to fit the spectra with a single power law as well. We take the frequency interval [0.1; 1.0] Hz, in which the power law is well characterized for all $T$ groups. The power-law indices become $\beta_{HF} = -1.91 \pm 0.02 \pm 0.09$, $\beta_{HF} = -2.06 \pm 0.03 \pm 0.02$, and $\beta_{HF} = -1.73 \pm 0.03 \pm 0.05$. These values are consistent with each other within $3\sigma$. The $\chi^2$ values can be found in Table \ref{tab:fittingparameters} and are comparable for the BPL and SPL fits.

\subsubsection{\label{subsubsec:redshift} Effect of GRB redshift}

\begin{figure}
        \centering
        \includegraphics[width=0.95\linewidth]{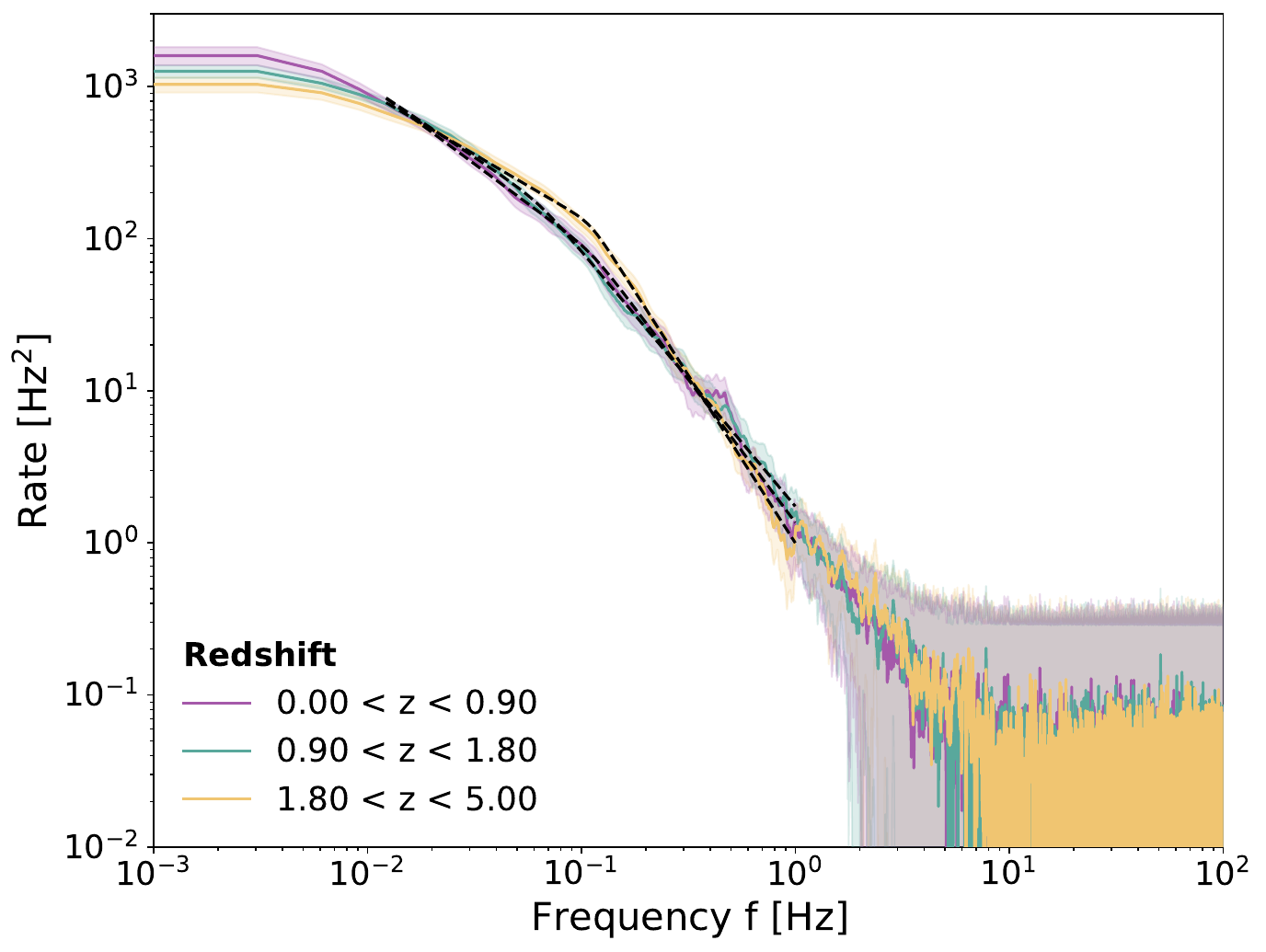}
        \caption{Average PDS for three redshift groups of LGRBs. The band around the PDS gives the $1\sigma$-error band. The dotted lines represent the best power-law fits.}
        \label{fig:combined_redshift}
    \end{figure}

    The measured redshifts of the bursts are used to correct the light curves for both the observed photon energies and arrival time intervals, resulting from the cosmic evolution over time. GRBs are known to spread out across a wide range of redshifts, with the highest photometric redshift estimated at $z \sim 9.4$ \cite{Cucchiara_2011}. In our sample, the redshift ranges from a minimum of $z = 0.0093$ to a maximum of $z = 8.0$. To analyze the redshift dependence of the power law, we divide the long GRBs based on redshift into bins of approximately 50 GRBs. Outliers with $z > 5.00$ are excluded, motivated by the redshift distribution shown in Fig. \ref{fig:grbdistribution}. This leaves a total of 158 long GRBs. The considered subgroups are $0.00 < z < 0.90$ (53 LGRBs), 0.90 $< z <$ 1.80 (54 LGRBs), and 1.80 $< z <$ 5.00 (51 LGRBs). The results are presented in Table \ref{tab:fittingparameters} and Fig. \ref{fig:combined_redshift}. The high-frequency power-law indices, ordered from low to high redshift, are $\beta_{HL} = -1.88 \pm 0.03 \pm 0.05$, $\beta_{HL} = -1.68 \pm 0.02 \pm 0.13$, and $\beta_{HL} = -2.20 \pm 0.02 \pm 0.07$. The high-redshift PDS tends to steepen more than the other spectra. The low- and mid-redshift power-law indices are consistent with each other within $2\sigma$. 

\subsubsection{\label{subsubsec:phases} Effect of GRB phases}

\begin{figure}
        \centering
        \includegraphics[width=0.95\linewidth]{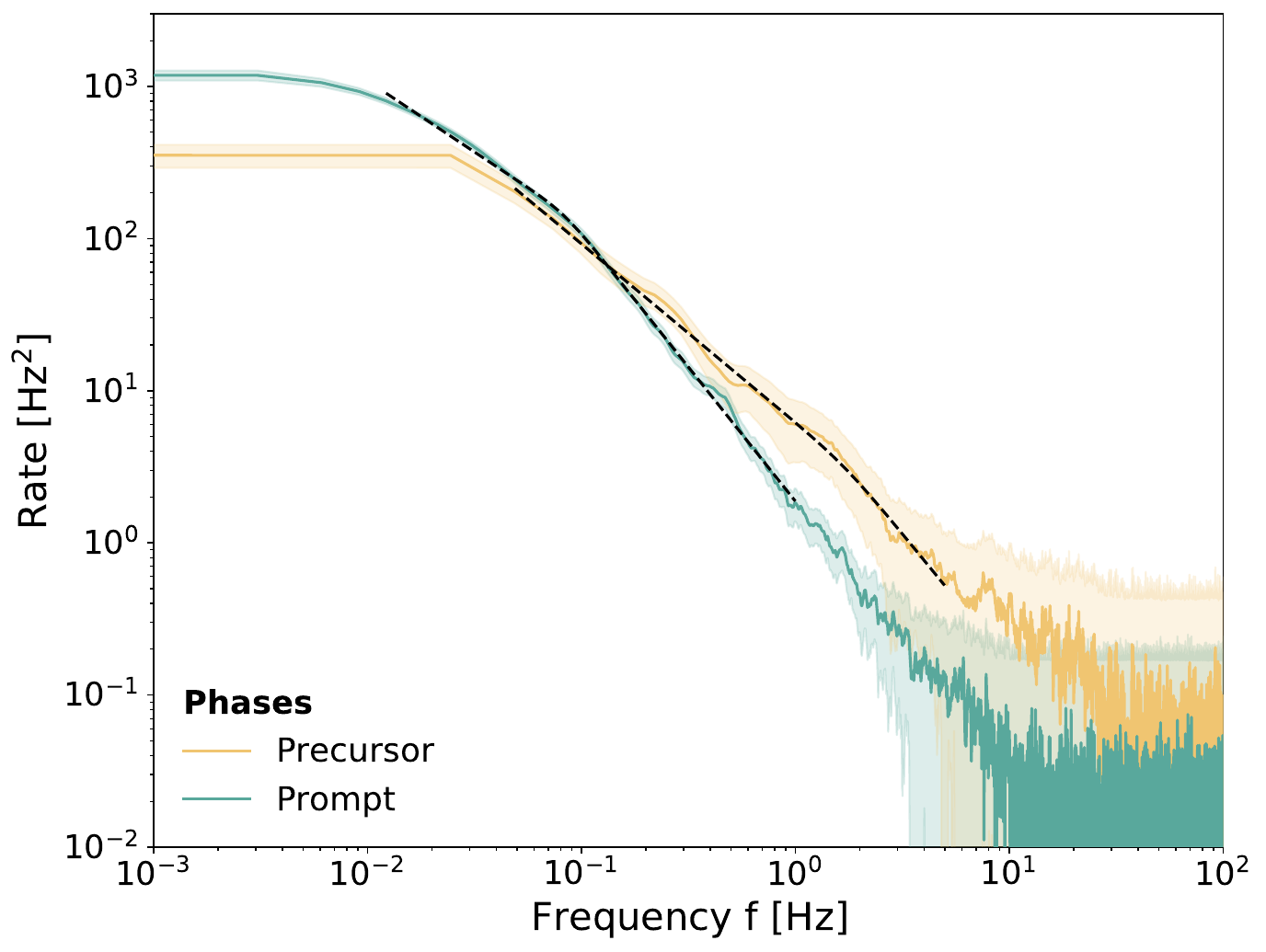}
        \caption{Average PDS for 26 precursor phases and 159 prompt phases of long GRBs. The band around the PDS gives the $1\sigma$-error band. The dotted lines represent the best power-law fits.}
        \label{fig:combined_phases}
    \end{figure}

    The phase-identification procedure allows us to divide the GRB light curve into distinct phases. Since the afterglow phase is not prominently observed in gamma rays with \textit{Fermi}-GBM, our analysis focuses solely on the precursor and prompt emission. Our sample includes 22 LGRBs with at least one precursor and the prompt phase of 159 LGRBs. Two GRBs have two precursors, while one GRB displays three. This yields in total 26 precursor phases and 159 prompt phases for constructing the average PDS of these two classes. 
    The durations of the precursor phases, identified in the redshift-corrected light curves, range from $T_{SB} = 0.033$ s to $T_{LB} = 37.35$ s, corresponding to features expected within the frequency range of $f_{LB} = 1/T_{LB} = 0.027$ Hz to $f_{SB} = 1/T_{SB} = 30.48$ Hz, with internal features extending to higher frequencies. For the prompt phase, durations span from $T_{SB} = 0.61$ s to $T_{LB} = 211.23$ s. The best-fit parameters are listed in Table \ref{tab:fittingparameters}, and the corresponding averaged spectra are shown in Fig. \ref{fig:combined_phases}. 
    For the prompt phase, the high-frequency power-law index is $\beta_{HF} = -1.76 \pm 0.01 \pm 0.04$, determined over the fitting interval $f = [0.01; 1.00]$ Hz. For the precursor phase, we extend the fitting interval to $f = [0.0268; 5.00]$ Hz, as the noise becomes dominant only beyond this range. The low- and high-frequency power-law indices are, respectively, $\beta_{LF} = -1.17 \pm 0.04 \pm 0.10$ and $\beta_{HF} = -1.7 \pm 0.2 \pm 0.2$. However, the large uncertainties in $\beta_{HF}$, the absence of a clear, visible break, and the relatively high-frequency break in the spectrum, at $f = 1.7 \pm 0.6 \pm 0.1$ Hz, suggest that the precursor PDS is better represented by a single power law. This is supported by a better $\chi^2/n_{d.o.f.}$ value for the single power-law model (increasing from $0.06$ to $0.11$). Then, the power-law index is $\beta_{HF} = -1.26 \pm 0.02 \pm 0.05$. This value aligns the $\beta_{LF}$ for the broken power-law model.  

\subsubsection{\label{subsubsec:noise} Effect of noise}

    To investigate the hypothesis that the power-law behavior and features are not inherent to the fluctuations of the GRB gamma-ray photons, we study the noise profiles of each of the above discussed categories. We expect that the slope of the noise PDS will reflect the duration distribution of the included light curves. The signal PDS, however, will additionally probe features inherent to the gamma-ray emission, and, therefore, exhibit a different slope. For each GRB, the duration of the phase emission is now taken in the noise region at the start or end of the time trace, as is shown in Fig. \ref{fig:lightcurve_example}. When this is not possible due to the limited duration of the time trace, we remove this GRB noise profile from the sample. This is done for every GRB population, and the exact same analysis is applied as before to construct the average PDS. Equation \eqref{eq:smoothbrokenpowerlaw} is then fitted to this average noise PDS. In the case of the $T$ groups, the precursor phase and short bursts, a single power law is used. The fitting occurs in the same fitting interval as for the signal regions. The results are compared to the results of the average signal PDS in Fig. \ref{fig:allresults} and discussed in Sec. \ref{subsubsec:noiseprofiles}. 

\subsection{\label{subsec:shortGRBs} Short GRBs}

\subsubsection{\label{subsubsec:redshiftcorrected} Redshift-corrected short GRBs}

\begin{figure}
        \centering
        \includegraphics[width=0.95\linewidth]{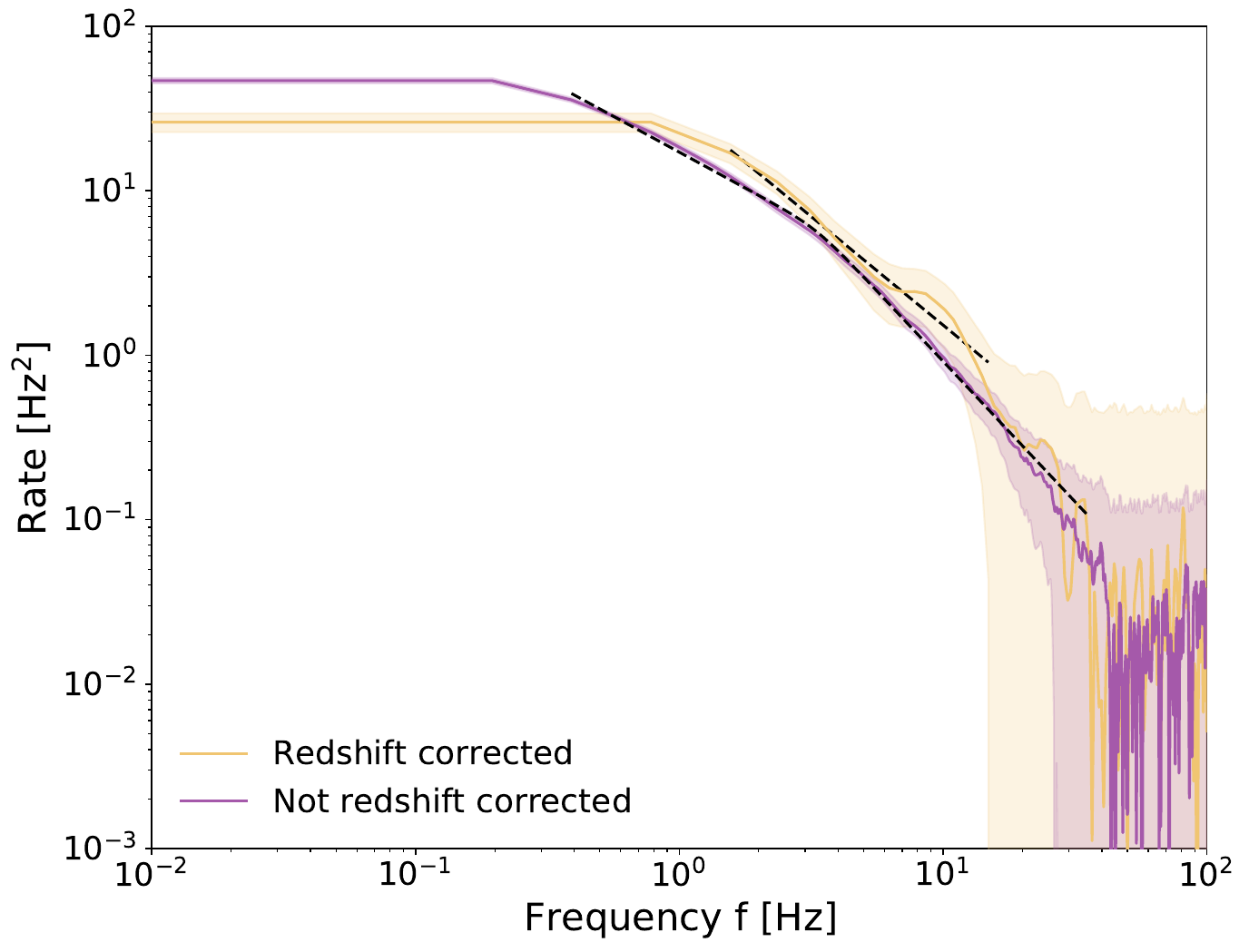}
        \caption{Average redshift-corrected PDS for 24 SGRBs and average non-redshift-corrected PDS for 395 SGRBs. The band around the PDS gives the $1\sigma$-error band. The dotted lines represent the best power-law fits.}
        \label{fig:combined_shortbursts}
    \end{figure}

    The dataset of 214 GRBs includes 27 SGRBs with known redshift. In three of these, our phase-identification procedure identifies emission zones with a duration of $T > 2$ s. Therefore, they are excluded from this analysis. The average PDS for the short GRBs is shown in Fig. \ref{fig:combined_shortbursts} with the best-fit parameters listed in Table \ref{tab:fittingparameters}. The power law in the spectrum is not very apparent. We believe this is due to the insufficient number of short GRBs. The shortest emission zone in our sample lasts $T_{SB} = 0.065$ s, while the maximal duration is $T_{LB} = 0.660$ s. That means we expect the features due to the variability between $f_{LB} = 1/T_{LB} = 1.516$ Hz and $f_{SB} = 1/T_{SB} = 15.404$ Hz. We decided to fit the single power-law model between those frequencies without extending to higher frequencies. This results in $\beta = -1.3 \pm 0.1 \pm 0.2$. This result should be taken with caution due to the small number of short GRBs. 

\subsubsection{\label{subsubsec:notredshiftcorrected} Short GRBs with unknown redshift}

    Complementary to the average redshift-corrected PDS, we investigate the average non-redshift-corrected PDS of short bursts to increase the statistics of the sample. As is shown in Fig. \ref{fig:grbdistribution}, the redshift distribution for short GRBs with known redshift ranges between $0.038$ and $2.211$. The redshift correction is, thus, minimal. We expect a similar redshift distribution for short bursts with unknown redshift, and, therefore, it remains reasonable to investigate the non-redshift-corrected PDS for short bursts. Between July 2008 and December 2023, the \textit{Fermi}-GBM telescope observed 602 short GRBs, including the 27 SGRBs with known redshift discussed above. We apply the same analysis procedures, outlined in Sec. \ref{sec:dataanalysis}, to these light curves to generate the average PDS. Within this dataset, the phase-identification procedure does not detect significant emission zones in the light curves of 171 short GRBs. This is likely due to the briefness of the bursts and the weakness of the peak, making their light curves unsuitable for the PDS analysis. Additionally, in 38 bursts, the phase-identification procedure identifies emission zones with a total duration exceeding 2 s, which we exclude from the sample. This results in a final selection of 395 short GRBs for constructing the average non-redshift-corrected PDS.
    
    The resulting average PDS is shown in Fig. \ref{fig:combined_shortbursts}. Since the temporal features emerge at higher frequencies, we fit the PDS starting from the frequency corresponding to the longest burst ($T_{LB} = 2.57$ s, or $f_{LB} = 0.39$ Hz) up to $f = 35$ Hz. This upper bound represents a compromise between the frequency associated with the shortest burst ($T_{SB} = 0.01$ s, and $f_{SB} = 94.35$ Hz), which falls within the noise-dominated region of the PDS, while maximizing the included PDS range. The corresponding best fit results in a high-frequency slope of $\beta_{HF} = - 1.71 \pm 0.03 \pm 0.07$.

\section{\label{sec:discussion} DISCUSSION}

\begin{figure*}
    \centering
        \includegraphics[width=0.9\linewidth]{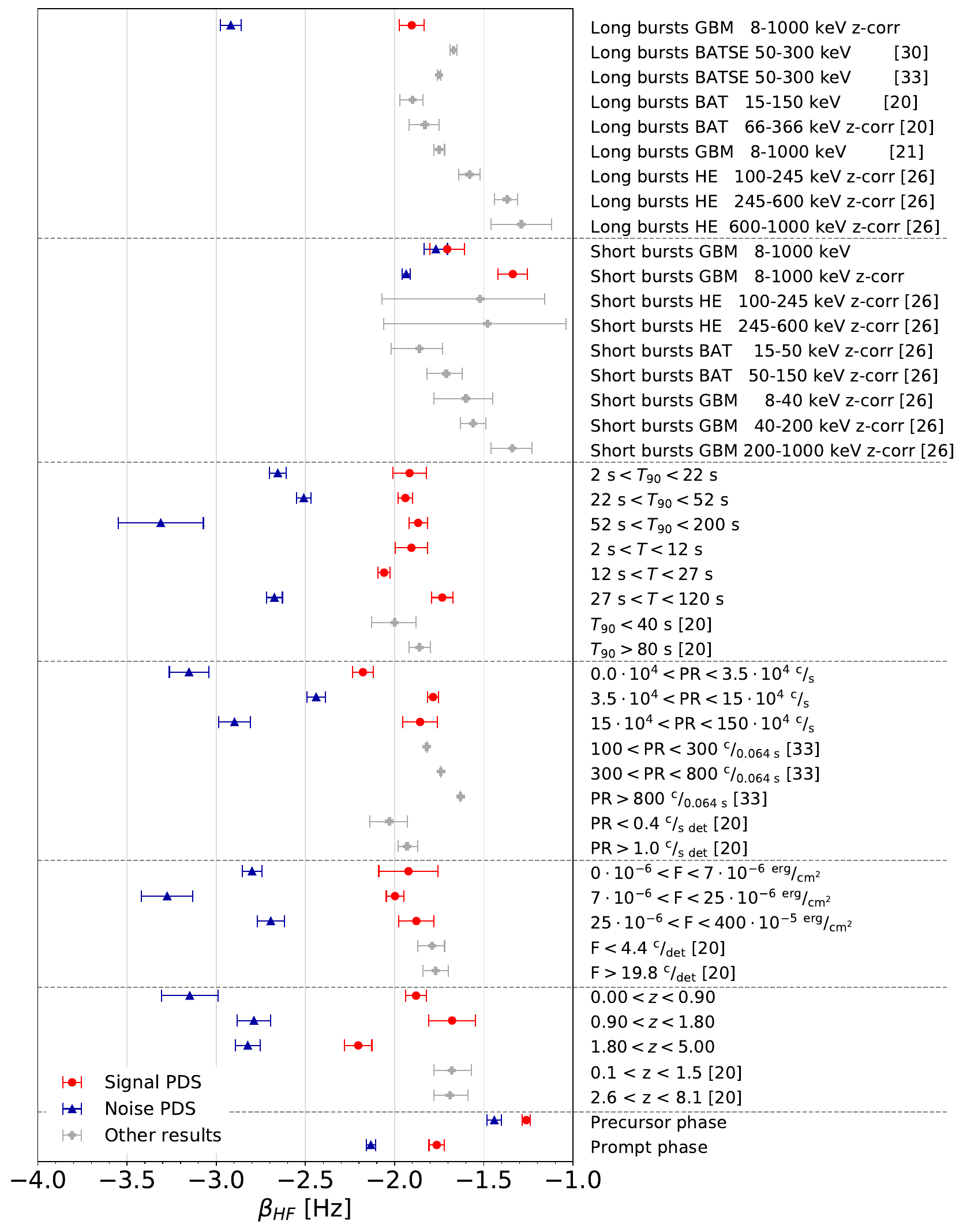}
        \caption{Representation of the evolution of the high-frequency slope $\beta_{HF}$ across all GRB groups. The red dots represent $\beta_{HF}$ for the signal phases; the blue triangles represent $\beta_{HF}$ for the noise phases. The error bars combine the statistical and systematic errors and represent the $1\sigma$ interval. If the spectrum is fitted with a single power law, this slope is shown. In addition, the gray crosses show relevant results from other studies. Abbreviations: SPL, single power law; OF, observer frame; RF, redshift-corrected frame.}
        \label{fig:allresults}
    \end{figure*}

    The high-frequency slopes for each of the different groups of GRBs are listed in Table \ref{tab:fittingparameters}. They are discussed below and compared to other studies in Fig. \ref{fig:allresults}. 

    \subsubsection{Long bursts} 

    Beloborodov, Stern, and Svensson \cite{Beloborodov_1998} analyzed 214 long and bright CGRO-BATSE GRBs ($T_{90} > 20$ s and PR $> 250$ counts per 64 ms bin) and found a best-fit slope $\beta_{HF} = -1.67 \pm 0.02$. In their follow-up paper \cite{Beloborodov_2000}, they extended the sample to include fainter bursts, resulting in a steeper slope of $\beta_{HF} = -1.75 \pm 0.01$. Guidorzi \textit{et al.}\ \cite{Guidorzi_2012} reported a high-frequency power-law slope of $-1.90^{+0.07}_{-0.06}$ for their full sample of long and bright BAT GRBs (without redshift correction). Within this dataset, 97 GRBs had a known redshift (\textit{z-silver sample}), yielding a corresponding power-law index of $-2.06^{+0.25}_{-0.18}$ (not redshift-corrected). For the \textit{z-golden sample} (49 GRBs with redshifts in the range $1.4 < z < 3.5$), the redshift-corrected power-law index was $-1.83^{+0.09}_{-0.08}$, showing no significant deviation from the uncorrected result of the same sample. Dichiara \textit{et al.}\  \cite{Dichiara_2013a} found a power-law index of $-1.75 \pm 0.03$ for their full sample of 205 \textit{Fermi} LGRBs using a different light-curve normalization (variance normalization). Zhou \textit{et al.}\ \cite{Zhou_2024} reported high-frequency slopes between $-1.29$ and $-1.58$ for long bursts in three different energy bands. Our result of $-1.90 \pm 0.01 \pm 0.07$ remains consistent with most findings within a 3$\sigma$ interval.   

    \subsubsection{GRB fluence and GRB peak rate} 

    Regarding the evolution of the high-frequency slope across different observed fluence bins, all results agree within a $2\sigma$ interval. No clear trend is observed in the low-frequency slope or break frequency. Guidorzi \textit{et al.}\ \cite{Guidorzi_2012} also found no trend with increasing fluence, reporting slopes of $-1.79^{+0.08}_{-0.07}$ and $-1.77^{+0.07}_{-0.07}$, although they do observe a steepening of the low-frequency slope.

    For the evolution with peak rate, the lowest peak-rate group exhibits a significant steeper power-law slope compared to the other two groups, with a 3$\sigma$-5$\sigma$ difference. The latter two groups remain consistent with each other within a $1\sigma$ interval. A direct comparison with other studies is challenging, as peak rate is often defined using different time binnings. Beloborodov, Stern, and Svensson \cite{Beloborodov_2000} identified three peak-count-rate bins (per 64 ms bin). They reported a flattening of the power law for brighter bursts, with a decrease of $\beta_{HF}$ from $-1.82$ to $-1.74$ to $-1.63$ as peak-count rate increases. This trend is consistent with our findings. Guidorzi \textit{et al.}\ \cite{Guidorzi_2012} found no significant difference between the high-frequency slopes of two peak-rate subsamples ($-2.03^{+0.11}_{-0.10}$ and $-1.93^{+0.05}_{-0.06}$), but they do report an increase by a factor of 5 in break frequency from the faint to the bright subset. In our analysis, we do not observe a significant trend in the low-frequency slope or break frequency. 
    
    \subsubsection{GRB duration}

    We investigated the evolution of the power-law indices with two types of duration: the observed $T_{90}$ and the $T$ parameter from the redshift-corrected light curves. For $T_{90}$, the results agree well within a $2\sigma$ interval and are consistent with those of Guidorzi \textit{et al.}\ \cite{Guidorzi_2012} ($-1.86^{+0.06}_{-0.06}$ and $-2.00^{+0.13}_{-0.12}$ for increasing duration). We notice that the low-frequency power-law index $\beta_{LF}$ decreases with increasing $T_{90}$. While $\beta_{LF}$ remains consistent within 2$\sigma$ for the two longest-duration groups, the difference compared to the shortest-duration group exceeds 5$\sigma$. This discrepancy might be linked to the break frequency $f_b$ occurring at a higher frequency for the shortest group. An artificial correlation between $f_b$, $\beta_{LF}$, and $\beta_{HF}$ may exist, as also suggested by Guidorzi \textit{et al.}\ \cite{Guidorzi_2012}: A higher $f_b$ tends to correspond to steeper values of $\beta_{HF}$ and $\beta_{LF}$. However, the break frequencies of the two longest-duration groups do not show a significant difference.  

    Since the observed $T_{90}$ parameter does not accurately reflect the true duration of the emission zones of the light curve in this specific energy band, duration-dependent evolution can be reliably analyzed only using $T$. The spectra within different $T$ bins, when fitted with the broken power-law model, yield high-frequency slopes and break frequencies that remain within a $2\sigma$ interval, while the low-frequency slopes are fully consistent within $1\sigma$. The results obtained using a single power-law fit exhibit lower systematic errors and remain consistent within $3\sigma$, aligning better with other findings in this analysis and prior studies.   
    
    \subsubsection{GRB redshift}

    No clear trend is observed in the evolution of the high-frequency slope with redshift. While the two lowest redshift groups agree within $2\sigma$, the highest redshift group deviates by more than $2\sigma$, but still less than $3\sigma$, from the lowest redshift groups. However, the low-frequency slopes suggest a possible trend: The slope for the highest redshift group differs from that of the lowest redshift group by more than $3\sigma$, although the mid-redshift group remains consistent with both.
    
    Guidorzi \textit{et al.}\ \cite{Guidorzi_2012} report no trend in the high-frequency slope for their peak-normalized light curves, with values of $-1.68 ^{+ 0.10} _{- 0.11}$ for $0.1 < z < 1.5$ and $-1.69 ^{+ 0.09} _{- 0.10}$ for $2.6 < z < 8.1$. However, for variance-normalized light curves, they do observe shallower indices at higher redshifts. They also find a hint of anticorrelation between observed fluence and redshift. In our sample, the correlation coefficients also suggest a weak anticorrelation between both parameters ($\rho = -0.17$, $r = -0.17$, and $\tau = -0.12$). Since the fluence-based analysis did not reveal significant deviations between different fluence groups, we believe that this relation does not substantially impact our results. A stronger correlation is found between redshift and peak rate, supported by all correlation metrics: $\rho = 0.42$, $r = 0.27$, and $\tau = 0.19$. This indicates that GRBs at higher redshifts tend to be observed with stronger pulses. This can be understood by noting that high-redshift GRBs are, in general, more luminous compared to their low-redshift counterparts, and this analysis considers the peak rate calculated from the redshift-corrected light curves. However, while the peak-rate analysis shows a flattening of the power law with increasing peak rate, we observe a steepening of the power law at higher redshifts. This relation is, thus, not able to fully explain the deviating results in both redshift and peak rate.
     
    \subsubsection{Precursor and prompt phase}

    The average precursor PDS, when fitted with a single power-law model, shows a significant deviation from the average prompt-phase PDS. Although the precursor phase appears consistent with the prompt phase when fitted using a broken power-law model, this consistency is largely driven by a local region of the PDS rather than the overall shape. 
    It is important to note that the precursor PDS is based on only 26 signal traces. Given the presence of many perturbations that influence the shape of the PDS, this limited sample size is likely insufficient to robustly characterize the temporal features of the PDS. Nevertheless, the significant difference in slope observed with the single power-law fit suggests that the precursor emission may arise from different processes or environments than those responsible for the prompt emission, constraining the origin of the precursor phase. This invites further investigation. 

    \subsubsection{Short bursts}

    The average PDS for short bursts in the redshift-corrected frame includes only 24 GRB light curves and does not exhibit a clear power-law behavior. As a result, this PDS is excluded from further analysis. In contrast, the average PDS of short bursts in the observer frame, constructed from 395 bursts, displays a clear power-law structure that is well described by a broken power-law model. The temporal features appear at significantly higher frequencies compared to long bursts, as evidenced by the location of the break frequency. 
    The low- and high-frequency slopes are consistent within $2\sigma$ with those found for long bursts. However, it is important to note that this PDS has not been redshift corrected. Nevertheless, Guidorzi \textit{et al.}\ \cite{Guidorzi_2012} report no significant differences between the redshift-corrected and uncorrected sample of 49 long bursts, suggesting that the observer-frame result includes meaningful information. We also compare our results with those from Zhou \textit{et al.}\ \cite{Zhou_2024}, who analyzed 11 bright Insight-HXMT-HE short bursts, 13 bright \textit{Swift} short bursts and 13 bright \textit{Fermi} short bursts without redshift correction across different energy bands. The results vary between $-1.34$ and $-1.86$, which are consistent with our findings. 
    
    \subsubsection{\label{subsubsec:noiseprofiles} Noise profiles}

    The results of the noise analysis are presented in Fig. \ref{fig:allresults}. For the long bursts, their respective subgroups, and the prompt phase, all signal spectra significantly deviate from their noise counterparts, indicating that the observed temporal features are not solely driven by the duration distribution of the GRBs in the sample. In contrast, the precursor phase behaves differently. Its noise result remains consistent with the signal within $3\sigma$, meaning we cannot exclude the possibility that the power law is not inherent to the signal photons. This observation further motivates deeper investigation into the precursor PDS, noting again that the precursor result is based on a small sample of GRBs.

    For the redshift-corrected short bursts, the noise result deviates from the signal result at the $5\sigma$ level. However, in the observer frame, the noise and signal PDS are not significantly different. These findings suggest that the duration distribution of short bursts may be the main factor in shaping the observed power-law behavior, in contrast to the scale-free processes that would rule in the long bursts. Since both emerge from a different progenitor, this hypothesis cannot be overlooked. The power law for long bursts is observed in individual spectra as well \cite{Guidorzi_2016, Dichiara_2016, Zhou_2024}. Zhou \textit{et al.}\ \cite{Zhou_2024} reported for their sample of 11 short bursts mean power-law indices comparable to the long bursts for the individual spectra.

    \subsubsection{Characteristic timescale}

    For individual time spectra, the break frequency is often associated to the duration of the dominant pulse in the light curve, referred to as the dominant or characteristic timescale $\tau \sim 1/2\pi f_b$ \cite{Lazzati_2002}. This definition is derived from the power-density spectrum of an exponential pulse with characteristic timescale $\tau$, which has a break at $f_b \sim 1/2\pi \tau$. Guidorzi, Dichiara, and Amati \cite{Guidorzi_2016} find that $\tau$ correlates with the $T_{90}$ for individual GRBs. They find timescales ranging from 0.2 to 30 s, with a logarithmic average of 4.1 s. Similarly, Zhou \textit{et al.}\ \cite{Zhou_2024} report values between 0.18 and 12.95 s (average of 1.58 s) for long bursts and between 0.01 and 0.08 s (average of 0.02 s) for short bursts, confirming the experimental relationship for both types. Although the average PDS provides only an approximate measure of the dominant timescale across a GRB sample, Guidorzi \textit{et al.}\ \cite{Guidorzi_2012} report characteristic timescales of approximately 5-8 s in observer frame, and 2-4 s in redshift-corrected frame. Dichiara \textit{et al.}\ \cite{Dichiara_2013a} observe a characteristic timescale of about $3$ s in the observer frame.

    In this analysis, all break frequencies (except those of the short bursts and precursor phases) cluster around approximately $f_b \sim 0.1$ Hz. This corresponds to a characteristic timescale of $\tau \sim 1.6$ s, which aligns with the redshift-corrected range reported by Guidorzi \textit{et al.}\ \cite{Guidorzi_2012}. When using the break frequency of the non-redshift-corrected bursts, we have a characteristic timescale of $\tau = 0.05$ s. This aligns with the results of Zhou \textit{et al.}\ \cite{Zhou_2024}. 

    \subsubsection{Theoretical models and simulations}

    For several GRB groups, the high-frequency power-law index remains consistent with the Kolmogorov $-5/3$ index, associated with the spectrum of velocity fluctuations in a turbulent medium. However, a trend toward a slope of $-1.9$ is observed, consistent with findings by Guidorzi \textit{et al.}\ \cite{Guidorzi_2012}. Notably, simulations by Zhang and Zhang using the  Internal-Collision-Induced Magnetic Reconnection and Turbulence (ICMART, \cite{Zhang_2011}) model yield power-law indices consistent with our findings and reproduce the typical variability of GRB light curves \cite{Zhang_2014a}. The power-law index of $-5/3$ can also be reproduced by the development of a turbulent dynamo in the jets, as described by Zhang, MacFayden, and Wang \cite{Zhang_2014b}, and the propagation of the jet through the stellar material, as shown by Morsony, Lazatti, and Begelman \cite{Morsony_2010}. Carballido and Lee investigated the time variability in the energy output as a result of turbulent motions in neutrino-cooled accretion disks \cite{Carballido_2011}. They simulated the effect on the power-density spectrum of the neutrino luminosity, which can serve as proxy for the variability of the central engine and relativistic outflow, and, thus, the GRB. They find a power law extending over multiple orders of magnitude with index $-1.7$ to $-2$, depending on the cooling mechanism. These values align well with our results. Other papers tried to explain the power-law features without invoking turbulence. Panaitescu, Spada, and Meszaros and Spada, Panaitescu, and Meszaros showed that the index of $-5/3$ can be obtained by adjusting the dynamics of the relativistic winds within the relativistic internal shock model \cite{Rees_1994, Panaitescu_1999, Spada_2000}, and Chang and Yi showed that the $-5/3$ slope can be recovered by adjusting the rise, decay and sampling timescales of the GRB pulses \cite{Chang_2000}. Finally, a recent paper by Loznikov argued that all features of the average PDS can be explained by light curves with two-sided pulses, governed by parameters as the duration and the shape of the pulses \cite{Loznikov_2024}.

\section{\label{sec:conclusion} CONCLUSION}

    For the first time, an extended analysis on the redshift-corrected average power-density spectrum has been performed. We studied 214 GRBs with known redshift, observed by the \textit{Fermi} Gamma-Ray Space Telescope between 2008 and 2023. This dataset includes 27 short GRBs and 187 long GRBs. Additionally, we examined 602 short GRBs without redshift correction to account for the limited statistics of short GRBs with known redshift. In most cases, the PDS is well described by a smooth broken power law. When a clear break was absent, a single power law was fitted over a restricted frequency range. The high-frequency power-law indices $\beta_{HF}$ generally cluster around $-1.9$, while the low-frequency indices $\beta_{LF}$ center near $-1.0$. The break frequency typically occurs around $0.1$ Hz for long bursts and occurs at $\sim 3.2$ Hz for short bursts. This suggests a characteristic timescale of $\tau \sim 1.6$ s for long bursts and $\tau \sim 0.05$ s for short bursts, potentially related to the durations of individual emission pulses in the GRB light curves. 

    The high-frequency index of the average PDS remains robust across the evolution of various parameters for long GRBs. In comparing different fluence groups, no significant deviations in $\beta_{HF}$ were observed. When analyzing PDS evolution by duration, we considered the true duration of the emission zones, as investigated using the $T$ value. These spectra are better fitted by single power laws, which are consistent with each other and with the overall $-1.9$ slope. Dividing GRBs by their $T_{90}$ reveals a decreasing trend in the low-frequency power-law index for longer GRBs. A slight discrepancy is noted between the lowest peak-rate group and the mid- and high-peak-rate groups. This is consistent with Beloborodov, Stern, and Svensson \cite{Beloborodov_2000}, which report a steepening of the high-frequency power law for dimmer bursts. However, this finding is not supported by Guidorzi \textit{et al.}\ \cite{Guidorzi_2012}, which examined only two peak-rate groups. A positive correlation between peak rate and redshift is present in our sample, though it does not fully explain the behavior of the low peak-rate group. Comparisons across redshift bins show largely consistent results, with a possible trend emerging in the low-frequency index.

    Because of insufficient statistics, the redshift-corrected average PDS of short GRBs does not allow strong conclusions about their power-law indices, necessitating further study. As a result, we also analyzed non-redshift-corrected light curves. The derived high-frequency index is consistent within $2\sigma$ with those of long bursts.

    For the first time, we examined the average PDS across different GRB light curve phases. Utilizing the method described in Coppin \textit{et al.}\ \cite{Coppin_2020}, we distinguished between the precursor, prompt, and noise phases of long GRB. The prompt phase yields a high-frequency power-law index of $-1.76 \pm 0.01 \pm 0.04$, consistent with the general $-1.9$. This is expected, as the prompt phase dominates the emission. The precursor phase is best fitted by a single power law, with a high-frequency index of $-1.262 \pm 0.023 \pm 0.005$. However, this group includes only 26 precursor phases, limiting statistical power. We, therefore, encourage further investigation into this phase. A significant difference between the precursor and prompt phase may indicate a distinct production mechanism or emission environment. The average PDSs of noise profiles from various groups reveal clear deviations from signal-phase results, with the exception of non-redshift-corrected short bursts and precursor phases. This suggests that the signal zones in these groups may arise from noise-related or duration-specific features rather than intrinsic gamma-ray emission.

    This study provides a foundation for understanding the average power-density spectra of redshift-corrected GRBs across various parameters and emission phases. Future work should focus on improving statistical power for underrepresented groups, such as precursor phases and short bursts with known redshift. Further simulations could clarify the physical origins of the observed power-law behaviors, in particular, the connection to turbulence and jet dynamics in GRBs. 
    
\begin{acknowledgments}
    We thank the referee for the valuable comments, improving the quality of this work. This research has made use of the public \textit{Fermi}/GBM data obtained through the High Energy Astrophysics Science Archive Research Center Online Service, provided by the NASA/Goddard Space Flight Center. We acknowledge the public catalog of GRBWeb. 

\end{acknowledgments}

\section*{DATA AVAILABILITY}

    The data that support the findings in this article are openly available at GRBWeb \cite{Coppin_2020} and the High Energy Astrophysics Science Archive Research Center Online Service \cite{fermidata}.

\appendix

\section{\label{app:backgroundcharacterisation} BACKGROUND CHARACTERIZATION}

    For the background characterization and phase identification, we adopt the methodology described in Coppin \textit{et al.}\ \cite{Coppin_2020}. As explained in Sec. \ref{subsec:dataprocessing}, we select the two or three NaI detectors that triggered on the burst. Their data are collected to construct the combined light curve. The GBM instrument collects three types of data files for each burst: CTIME (continuous time), TTE (time-tagged event), and CSPEC (continuous spectroscopy) data. While the TTE data are used for the light curve itself owing to its fine temporal resolution, the CTIME data files, which provide binned photon counts with 256-ms resolution across eight energy bins \cite{Paciesas_2012}, are preferred to estimate the background rate. They allow for a stable and robust fit over a long time window, available from $-1000$ s before the trigger time of the GRB. Below, we summarize the method. Further details can be found in Ref. \cite{Coppin_2020}. 
    
    The observational window of the CTIME data spans 2000 s centered on the trigger time $T_0$ for each detector per burst. Background fits are created for the light curves of all selected individual detectors and the combined histogram. This process starts at the earliest data point, $t = t_0$. A linear fit is performed over the interval [$t_0$, $t_0$ + 25 s] to approximate the rate in this region. This linear fit is then used to predict the rate at $t = t_0 + 35$ s. The procedure is repeated gradually, by advancing one second to produce a predicted rate profile $r_{pr}$ over the entire light curve. Subsequently, the averaged rate $r_{av}$ is computed for each time step $t$ over a symmetric time interval of 2.5 s around $t$. The predicted rate $r_{pr}$ and averaged rate $r_{av}$ are then compared using the following criteria:

    \begin{align}
        r_{th} = 3 \cdot \sqrt{\frac{ r_{pr}}{2.5}}, \label{eq:r_th} \\
        r_{rms} = 1.5 \cdot \sqrt{\frac{r_{pr}}{2.5}} \label{eq:r_rms}.
    \end{align}

    If the average rate shows a 3$\sigma$ excess based on Poisson statistics, compared to the predicted rate, i.e. ($r_{av} - r_{pr}) < r_{th}$, the region is classified as background. If this condition is not met, the analysis progresses by 30 s, marking this region as a potential emission zone. For such regions, a secondary check is performed on the last 10 s of the potential emission zone. If the root mean square (rms) of ($r_{av} - r_{pr}$) lies within the 1.5$\sigma$ Poisson interval defined by Eq. \eqref{eq:r_rms}, these 10 s are classified as background dominated. Otherwise, the analysis continues by advancing another 30 s, repeating the above steps until the entire light curve is evaluated.

    \section{\label{app:phaseidentification} IDENTIFYING THE EMISSION PERIODS}

    The identification of the signal regions is based on a Bayesian block treatment of the redshift-corrected TTE light curves \cite{Coppin_2020, Scargle_2013}. They are subsequently used to compute the power-density spectrum, as explained in Sec. \ref{sec:dataanalysis}. The procedure for identifying the emission periods relies on the Bayesian blocks of both the combined and individual redshift-corrected TTE light curves, along with the corresponding background levels, which we correct for redshift. The Bayesian block procedure, as described in Ref. \cite{Scargle_2013}, generates histograms with variable bin widths, sensitive to significant variations in the light curve rates. This allows for a first discrimination between signal excess and background noise. 
    
    The background rates, estimated using the procedure in Appendix \ref{app:backgroundcharacterisation}, are subtracted from the Bayesian block rates. A region is classified as signal if the signal excess exceeds a predefined threshold value. This threshold, determined in Ref. \cite{Coppin_2020} through an optimization procedure, was initially established for \textit{Fermi}-GRBs without redshift correction and set at 30 Hz. Since this analysis includes redshift corrections, the threshold value is adjusted for each GRB light curve to $(1+z) \cdot 30$ Hz. Additionally, any candidate signal region must exceed a minimum duration of 5 ms, consistent with the binning resolution adopted for the binned TTE light curves. 
    
    Signal regions are identified independently in the combined and individual light curves. The latter are used to validate the signal regions in the combined histogram. In order to legitimate a signal region in the combined histogram, it must overlap with a signal region in at least two individual light curves. This ensures that statistical fluctuations in individual light curves do not propagate into the combined histogram. Furthermore, we impose a criterion on the quiescent period between two emission periods: The relative Poisson uncertainty of the intermediate period must be less than 5\%. Consequently, quiescent periods shorter than $\sim$ 0.2 s are excluded.  
    
    If only one emission zone is identified, it is classified as the prompt emission. In cases with multiple emission zones, the period with the highest photon count is designated as the prompt burst. Emission zones occurring before the prompt phase are classified as precursors, while those occurring after the prompt phase are designated as afterglow phases. 

    In ten GRBs, our procedure fails to identify any photon count excess above background level, resulting in no detected emission zones. These GRBs are GRB 090916009, GRB 100413732, GRB 110106893, GRB 110818860, GRB 120907017, GRB 140829880, GRB 140906175, GRB 150727793, GRB 161017745, and GRB 171222684. In three GRBs, we were not satisfied with the phase identification. Therefore, we modified the emission zones. Their original and modified phase identification are listed in Table \ref{tab:modifiedphaseidentification}.

    \begin{table}
        \centering
        \caption{Original phase identification and modified phase identification for GRB 140508128, GRB 141221338, and GRB 161117066. }
        \label{tab:modifiedphaseidentification}
        \begin{tabular}{|p{2.4cm}|p{2.7cm}|p{2.7cm}|}
        \hline 
            Name GRB & Original phase & Modified phase \\
            \hline 
            \hline 
            GRB 140508128 & [-0.41 s; 5.69 s], [12.05 s; 13.98 s], [19.11 s; 20.47 s] & [-0.41 s; 29.91 s] \\
            \hline 
            GRB 141221338 & [-0.44 s; 1.67 s], [5.04 s; 6.00 s], [97.70 s; 114.63 s], [180.33 s; 194.14 s]  & [-0.44 s; 1.67 s], [5.04 s; 6.00 s], [97.70 s; 140.81 s], [180.33 s; 194.14 s] \\
            \hline 
            GRB 161117066 & [-0.26 s; 26.47 s], [40.34 s; 55.74 s] & [-0.26 s; 55.74 s] \\
            \hline 
        \end{tabular}
    \end{table}

\nocite{*}

\bibliography{apssamp}

\end{document}